\documentclass[nonacm]{acmart}
\setcopyright{none}
\renewcommand\footnotetextcopyrightpermission[1]{} 
\newcommand{\hidegraphics}[1]{}
\renewcommand{\hidegraphics}[1]{#1}
\def\PDT{PDT}
\def\PDTs{PDTs}
\newcommand{\cut}[1]{}
\newcommand{\tr}[2]{#1}

\AtBeginDocument{%
  \providecommand\BibTeX{{%
    \normalfont B\kern-0.5em{\scshape i\kern-0.25em b}\kern-0.8em\TeX}}}

\usepackage{epsfig,url,subcaption,enumitem,lipsum}
\usepackage{algorithm}
\usepackage{amsmath}
\usepackage[noend]{algpseudocode}
\usepackage{listings}
\usepackage{booktabs} 
\usepackage[skip=3pt]{caption}

\newcommand{\wfiveh}[0]{{\sc w5h}}
\newlength{\tablength}
\newlength{\spacelength}
\newcommand{\tab}{\hspace{\tablength}}

\title{Supporting Human Memory by Reconstructing Personal Episodic Narratives from Digital Traces}

\begin{document}

\author{Varvara Kalokyri }
\affiliation{Department of Computer Science, Rutgers University}
\email{v.kalokyri@cs.rutgers.edu}

\author{Alexander Borgida }
\affiliation{Department of Computer Science, Rutgers University}
\email{borgida@cs.rutgers.edu}

\author{Am\'elie Marian }
\affiliation{Department of Computer Science, Rutgers University}
\email{amelie@cs.rutgers.edu}


\begin{abstract}
Numerous applications capture in digital form aspects of people’s lives. The resulting data, which we call Personal Digital Traces — \PDT{}s, can be used to help reconstruct people’s episodic memories and connect to their past personal events. This reconstruction has several applications, from helping patients with neurodegenerative diseases recall past events to gathering clues from multiple sources to identify recent contacts and places visited – a critical new application for the current health crisis. This paper takes steps towards integrating, connecting and summarizing the heterogeneous collection of data into episodic narratives using scripts--- prototypical plans for everyday activities. Specifically, we propose a matching algorithm that groups several digital traces from many different sources into script instances (episodes), and we provide a technique for ranking the likelihood of candidate episodes.
We report on the results of a study based on the personal data of real users, which gives evidence that our episode reconstruction technique  1) successfully integrates and combines  traces from different sources into coherent episodes, and 2) augments users’ memory of their past actions.
\end{abstract}

\maketitle
\section{Introduction}
Memory has a fundamental role in life and is critical to our everyday functioning. We use  memories to maintain our personal identity, to support our relationships, to learn, and to solve problems. Various new technologies and applications make it possible to digitally capture a huge amount of personal data about aspects of our lives, such as digital communications with friends, events we participate in, trips we make. Our actions result in a multitude of \PDT{}s, kept in various locations in the cloud or on local devices: messaging and emails, calendars, location check-ins (e.g., Facebook Places, GPS tracker), online reservations (e.g. Opentable, Ticketmaster), reviews (e.g, Tripadvisor, Yelp), purchase history (e.g. credit card statements), photos, etc. 
Like a ``memex'', as envisioned by Vannevar Bush~\cite{memex}, this personal data collection can significantly help users to remember everyday situations. For example, users could use connections in their data to quickly retrieve digital artifacts, such as minutes from a specific meeting, pictures they took at a birthday party, the expense report for a specific trip they went on, or to help them recall specific memories, such as the names or even faces of people they  interacted with, perhaps years ago.

Furthermore, the current health crisis has highlighted the need for discovering the whereabouts and interactions of people to enable contact tracing. Specifically, contact tracing is an investigative, often laborious process, that involves public health officials interviewing a subject to reconstruct their activities to identify everyone they have been in contact with \cite{eames2003contact, kiss2005disease, huerta2002contact, world2015implementation}\footnote{Much has been reported recently about ``digital contact tracing systems'' \cite{GoogleApple, raskar2020apps, PEPP-PT}. These  are really exposure alerting systems that allow users to be notified if they were potentially in contact with an infected individual. 
Exposure alerting and traditional contact tracing go hand in hand in the fight to decrease the spread of the disease.}.
Traditional contact tracing is time-consuming and depends entirely on users' memories of events. However, there is evidence \cite{bradburn1987answering} that humans
are less likely to encode and find it harder to retrieve {\em routine} experiences. 
This makes it hard for people to recall information about where and with whom they have been a while back (which, in case of COVID-19, can be up to 2 weeks ago). However, information about this is being continually recorded (actively or passively) by their digital devices,  and could be leveraged to help users recall their past actions. For example, a system could help users reconstruct from \PDT{}s the friends they have eaten dinner with, the times they went to a grocery store, and the transportation they took. 

 There has been extensive research in the area of life-logging, pioneered by Bell~\cite{mylifebits}, where the vision is to enable "total recall" of our lives through "total capture" of personally relevant information \cite{blanchette,czerwinski2006digital}. Such information includes the digital traces we work with: email, instant messages sent and received, web sites visited, credit-card transactions etc., as well as other data, such as images, video, and location data. This vision has its detractors, such as Sellen and Whittaker~\cite{sellen2010beyond}, who argue that rather than storing a complete lifelog,  systems should focus on selectively identifying effective retrieval cues to jog users' memories, and that life-logging systems should not replace human memory but rather support it, with the ultimate goal of deriving meaning from the collected data.

This work takes steps towards supporting human memory through episodic narratives, an idea based on psychology and cognitive science.  More specifically, the literature on the psychology of human memory indicates that people have two different kinds of memory: ``semantic'' and ``episodic'' memory \cite{tulving1972organisation,conway1993structure}.  Semantic memory refers to general knowledge of the world (e.g., you have to pay when buying something) whereas episodic memory refers to the capacity to re-experience specific episodes from the past (e.g., the occasion you went out to dinner to celebrate your 40th birthday).
 
 We combine representation of both semantic and episodic  memory in order to help users recall information of  past events. Our approach aims to organize \PDT{}s into episodes, while automatically extracting information about the relevant people and context. 
  This organization will allow  creating a personal knowledge base, which users may query to remember a particular event. In addition, our approach aims to provide users with a narrative, 
  which can subsequently be viewed to stimulate memory; this could be particularly useful in a variety of situations, such as patients with memory difficulties. We believe that our approach will help users recollect aspects of past everyday experiences that have subsequently been forgotten, and thereby form a powerful retrospective memory aid.

In this paper, we propose an approach to integrate and connect \PDT{}s from a variety of 
sources into coherent episodic narratives in order to support human memory. Our approach is centered on the use of so-called \textit{scripts}, first introduced by Schank and Abelson \cite{schank1977}. Script definitions model dynamic aspects of semantic memory. The paper reviews our conceptual model for describing entities  (including \PDT{}s) and scripts, and presents a matching algorithm that groups heterogeneous \PDT{}s into candidate script instances ({\em episodes}). The \PDT{} types include emails sent/received, posts on social media, bank transactions, calendar entries, location data, phone text messages and photos. Such information can be found in a variety of sources, often extracted through service APIs, but also in files.\footnote{We immediately acknowledge the sensitive nature of this information, and the very important privacy issues that they raise. In our current work, all information obtained resides on an individual’s own mobile phone. And users just give Yes/No answers in experiments, without disclosing personal information.}
An important component is a scoring technique for ranking the candidate episodes based on the strength of the evidence PDTs provide. This is made necessary  by the prototypical/heuristic nature of scripts, and the sparseness of the evidence for them. Finally, we report on the results of our approach to episode reconstruction  for the {\tt Eating\_Out} script based on personal data of real users, which shows that our approach can successfully integrate and connect \PDT{}s into script instances, and augments users' memory of their past actions, even if these happened less than a month ago.

\section{Personal Data Integration}
\label{sec:integration}

One of the main challenges in integrating \PDT{}s lies in the fact that data is scattered over many disparate sources, with different data models. To overcome this fragmentation and heterogeneity, one needs a formal conceptual model to represent personal data. 

 According to the Cognitive Science and Psychology literature, a natural way to remember past events is by any pertinent {\em contextual} information, 
which includes answers to the ``what, who, where, when, why, how'' (\wfiveh) questions \cite{schacter2002seven}. 
For example, if you try to remember the name of a restaurant you visited, questions like ``When did I go to that restaurant?'' and ``Who was with me?''   will be helpful \cite{jones2007personal}. 

Our prior work \cite{iiweb} on modeling personal data defined the \wfiveh{} model, 
which interpreted the six contextual dimensions as follows: What (content), Who (with/from/to whom,...), Where (physical or logical), When (time and date, but also what was happening concurrently), Why (goals, and sequences of events that are assumed to be causally connected), How (application, author, environment).

 \begin{figure}
\centering
\includegraphics[width=0.8\textwidth]{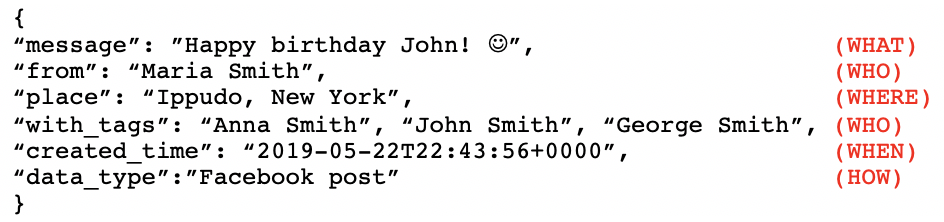}
\caption{Simplified Facebook post analyzed according to the \wfiveh{} model}
\label{fig-fbPost}
\end{figure}

Figure~\ref{fig-fbPost} presents a \PDT{}  from a Facebook post, with each piece of information identified as belonging to one of the six proposed dimensions. 
 Information from an e-mail, say, would also be represented using \wfiveh{} (e.g., {\tt from}, {\tt to}, {\tt cc}, {\tt bcc} would be part of the Who for the email). In both cases, this information can be represented in an ontology-language like OWL using entities and properties both organized in subsumption hierarchies. For example, {\tt from} and {\tt with\_tags} would be sub-properties of {\tt who} for Fb posts. 
The mapping from each data source to sub-properties of \wfiveh{} is performed by code requiring an understanding of that service's API.

\begin{figure}
\centering
\includegraphics[width=0.8\textwidth]{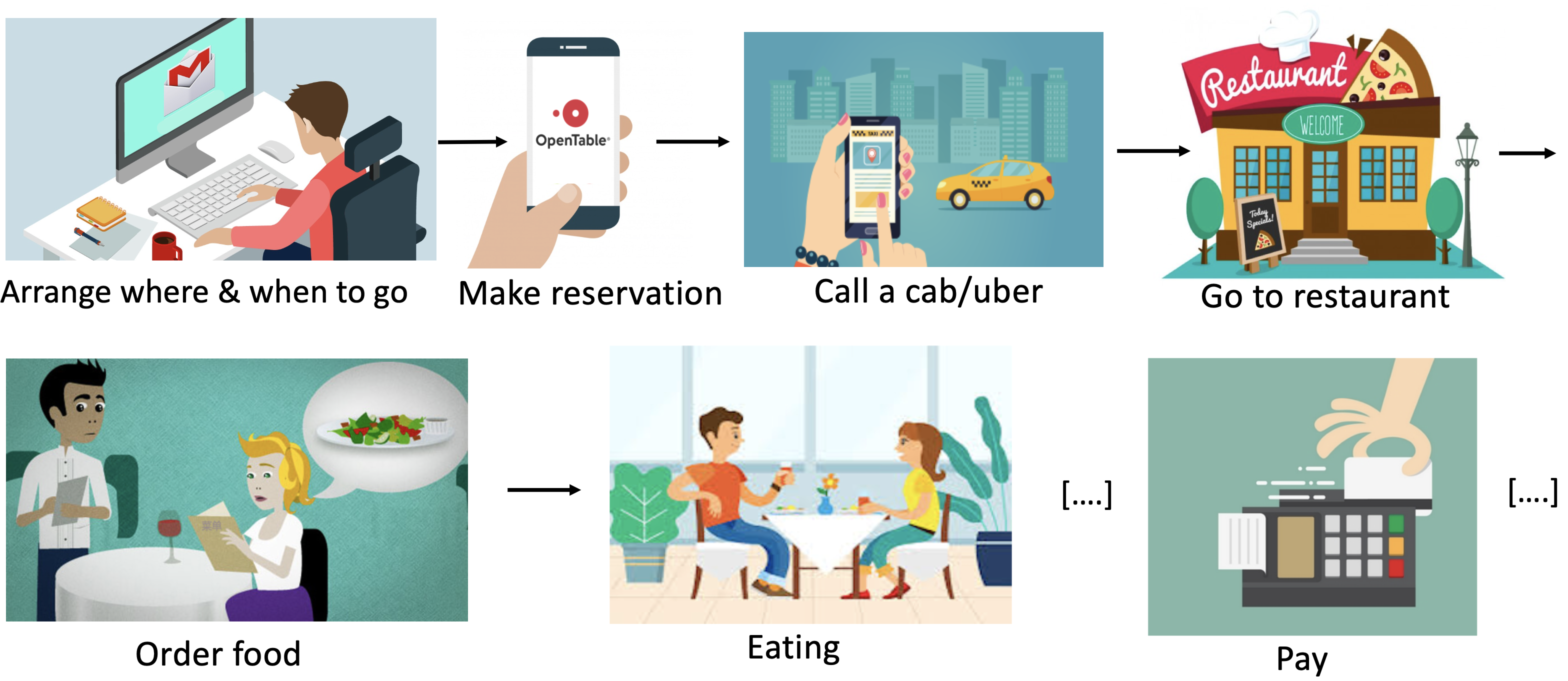}
\caption{Eating\_Out script with possible event flows}
\label{fig-script}
\end{figure}

 In addition, since  our work focuses on enhancing users' memory through the events in their lives, it is also necessary to provide a conceptual model for events, both atomic and complex ones. For this purpose we adopted the model proposed in \cite{usOdbase}.
 Recall that our goal for using scripts is to organize  \PDT{}s, abstract out relevant information, and help humans remember their events. An example of a script would be ``Going out to eat at a restaurant''. This script would provide a description of possible ``event flows'' as shown in Figure~\ref{fig-script}.
 The kinds of steps (atomic actions or sub-scripts) comprising a script, and their partial order, are knowledge we acquired throughout our lives; e,g, the fact that eating out requires, among others, organizing the outing, possibly making a reservation, getting to the restaurant, paying for the meal, etc. Each of these actions leaves a \PDT{}, which provides evidence for that particular event. For example, organizing the outing (e.g., via emails/text messages), possibly making a reservation (e.g, via OpenTable, which confirms by email), getting to the restaurant (e.g., using Uber/taxi, which leaves both a payment trace and an email confirmation), paying for the meal (e.g., with a credit card), etc. However, not every step occurs  every time one goes out to eat  (e.g., some restaurants do not require or even accept reservations), the order of some steps is not sequential (pay before or after eating?), and even more importantly we may find no digital evidence of some steps (e.g., when paying by cash).
 Therefore such scripts are partially instantiated, and should be scored based on the evidence for them.  This applies to any type of script that the system is told about: ``Shopping at the supermarket'', ``Visiting a doctor'', ``Going on a trip'', etc.

Scripts have properties describing: (i) their goal (for purposes of human explanation); (ii) summary information of the participants in the plan, as well as other descriptive properties, especially \wfiveh{} aspects; (iii) component sub-scripts and atomic actions; (iv) information about (dis)allowed sequencing and timing of sub-scripts/atomic actions. Items (iii) and (iv)
describe how the script achieves its goal, and make scripts resemble prototypical workflows.
 
In the next section, we describe the algorithm for creating episodes (instances of scripts) that a user may have been involved in, based on the \PDT{}s in the user's database. 

\section{Script Instantiation from \PDT{}s}
\label{sec:algo}

We gather \PDT{}s from various sources and use them as ``(noisy) sensors'' indicating the possible occurrence of corresponding (sub)scripts/ atomic actions. We use our extraction tool proposed in \cite{kalokyri2018yourdigitalself} to identify and retrieve data from currently popular services and sources of \PDT{}s. The resulting data collection is stored on the user's personal device, where there is no access from external source or service; this helps maintaining data privacy.

\subsection{Algorithm}
Our algorithm starts with the scripts we are interested in recognizing, and a large database of \PDT{}s (the majority unrelated to any script!). Our goal is to create candidate episodes (instances of scripts) that the user may have been involved in, and relate them to the \PDT{}s we have.

Consider a script that a person has been engaged in, for example, Eating\_Out. This script has many events  (e.g. invite people, discuss the specifics, make a reservation, go the place, pay the bill, post photos, etc... ). Some of these events provide strong evidence that the person was actually engaged in this script episode.
 For example, a payment to a restaurant is a strong evidence that the person actually went to it. In contrast, an email mentioning ``dinner'' or ``lunch'' is much weaker evidence for planning to go out to eat, and in turn this activity is weaker evidence for having gone out, since the planning may not have been completed, or the event cancelled. Our algorithm uses this idea of strong and weak evidence to rank candidate instances of   scripts.

\begin{algorithm*}
\caption{Algorithm for constructing instances of script S}
\label{algo}
\begin{algorithmic}[1]

\State $S := $ script definition 
\State $D  :=$  \PDT{}s indicating any potential instance of script S;
\State $Candidates = \{\}; $
\For {each d $\in$ D } \{
\State $ c_d := $ new instance of script S, based on d;
	\State  $c_d.score :=$ assign score based on strength of evidence;
	\State  $c_d.w5h :=$ extract \wfiveh{} information from d and add it to $c_d$;
\State $Candidates.add(c_d)$
\EndFor
\State\}
\State \textbf{repeat} until no changes in $Candidates$  \{
\State \tab $ MergeSet :=$  \{ $c_d$ in Candidates such that there is sufficient corroboration that they refer
to the same real-world episode \}; 
\State \tab $Candidates$ := (Candidates $-$ MergeSet) $\cup$ \{$c_d'$:=combine(MergeSet)\};
\State \tab $c_d'.score :=$ update score based on new strength of evidence;
	\State \tab $c_d'.w5h :=$ update \wfiveh{} information and add it to $c_d'$;
  \State \tab Use details of script $S$ to look for additional \PDT{}s $d$ that
       could be relevant to instances in $Candidates$;\\
   \}
\end{algorithmic}
\end{algorithm*}

\subsubsection{Script Syntax}
 
\begin{figure}
\centering
\includegraphics[width=0.8\textwidth]{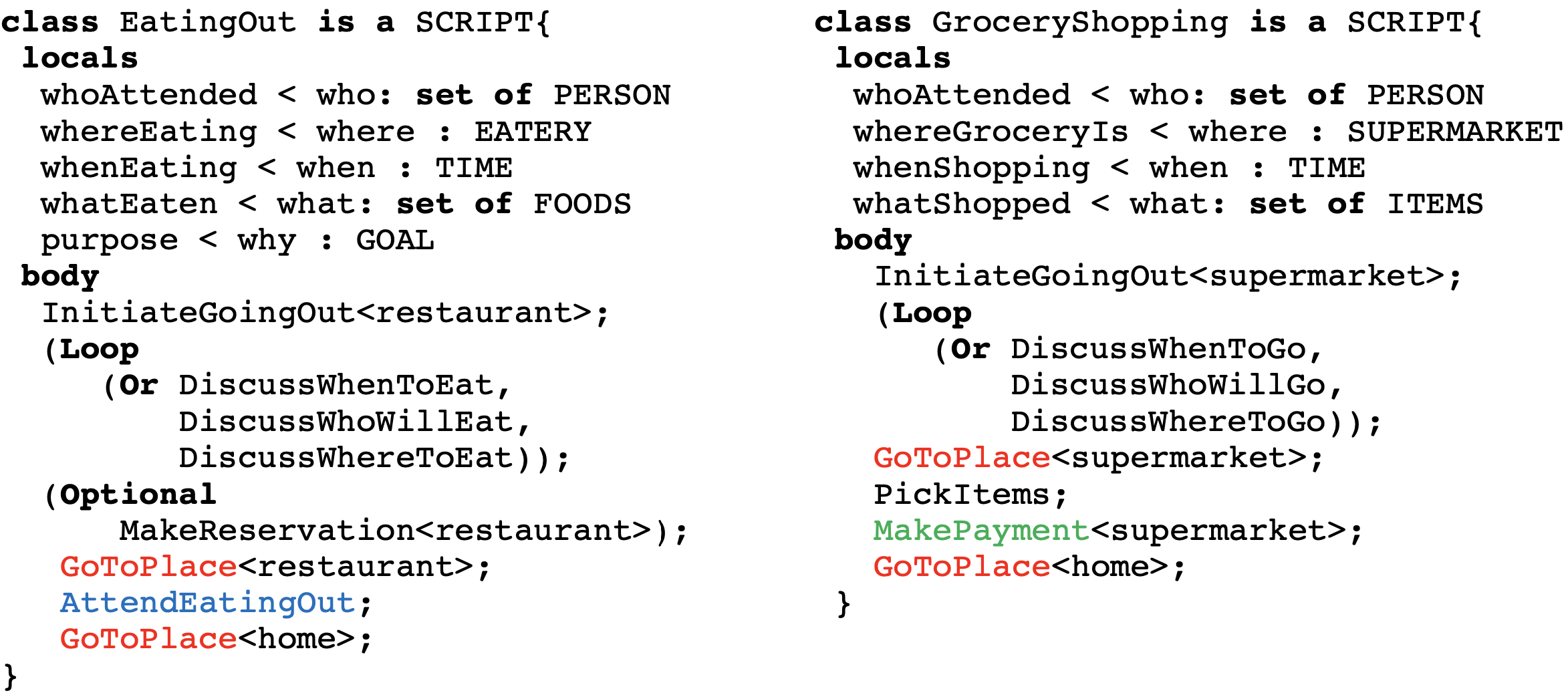}          
\caption{Definition of EatingOut and GroceryShopping scripts}
\label{fig-scriptDef}
\end{figure}

\begin{figure}
\centering
\includegraphics[width=0.8\textwidth]{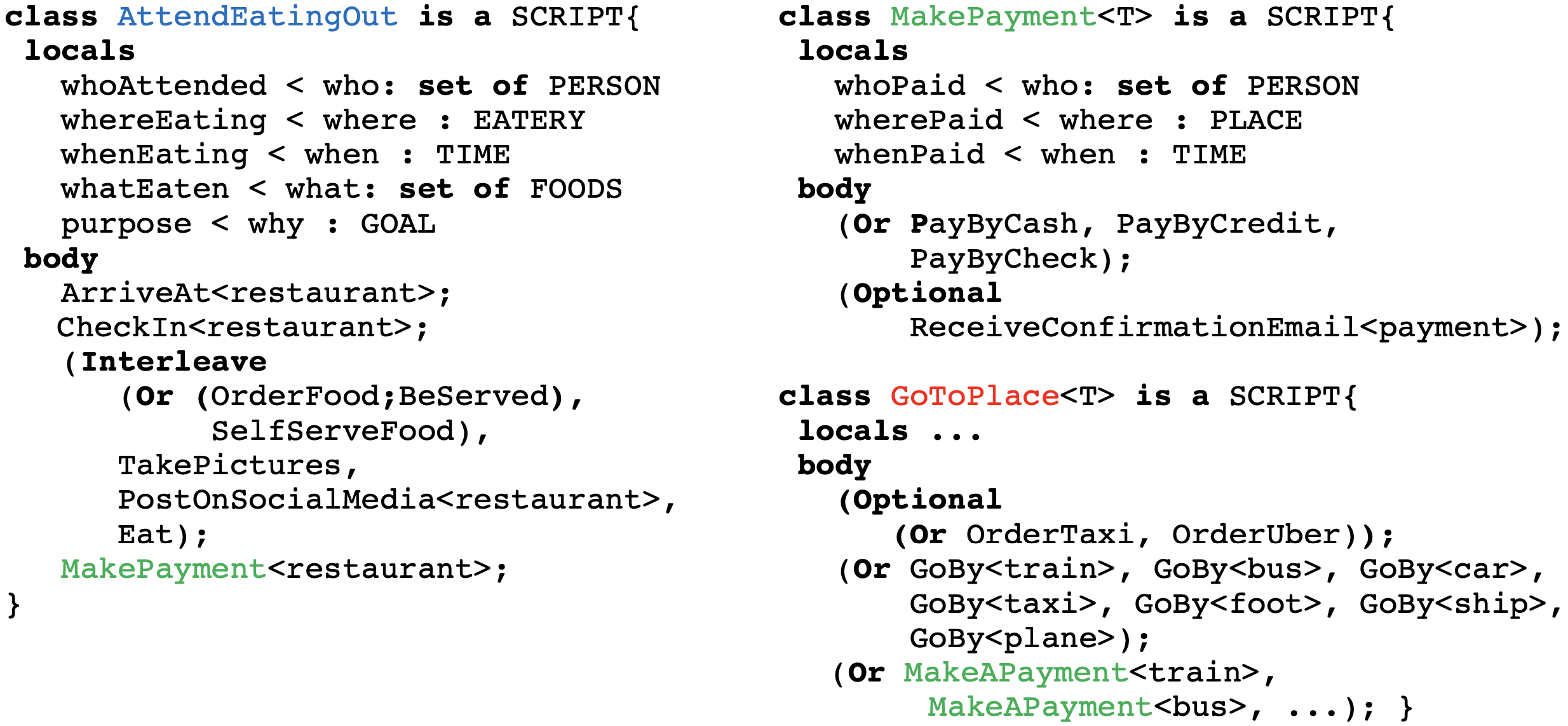}          
\caption{Definitions of the AttendEatingOut sub-script and the parametric MakeAPayment<T>, GoToPlace<T> sub-scripts.}
\label{fig-sub-scriptDef}
\end{figure}

A script specification consists of a top-level (outer) script that we want to instantiate (e.g. Eating\_Out script), several sub-scripts and atomic actions, as well as the sequencing relationship among them. In addition, every top-level script and sub-script should contain its \wfiveh{} properties. Figure~\ref{fig-scriptDef} shows examples of the Eating\_Out and the Grocery\_shopping script definitions. Both have local \wfiveh{} properties, and a body of atomic actions and sub-scripts (colored ones).  Figure~\ref{fig-sub-scriptDef} shows some of their sub-script definitions: AttendEatingOut, MakeAPayment$<$T$>$, and GoToPlace$<$T$>$. 
All (sub-)scripts and atomic actions have their own \wfiveh{} properties declared in their definition which need to be parsed as well. For instance, the Eating\_Out script has \wfiveh{} properties like whoAttended, whereEatingOccured, and whenEatingOccured.

 Note that some sub-scripts, such as MakePayment and GoToPlace, are parametric/generic, and are used in multiple places.
The argument replacing the formal parameter can be used as in procedures, searching text for a string for example, or the script is written as a case-statement, based on the  parameter value; this offers a way to organize cleanly the code for script recognition from \PDTs.


\subsubsection{Retrieving document set D indicating script instantiation}

After parsing the script, the next step is to find the set D of documents that provide evidence that a script instance has taken place. This corresponds to finding all atomic actions/sub-scripts for which there is strong evidence of having occurred. 

\begin{figure}
\centering
\includegraphics[width=0.6\textwidth]{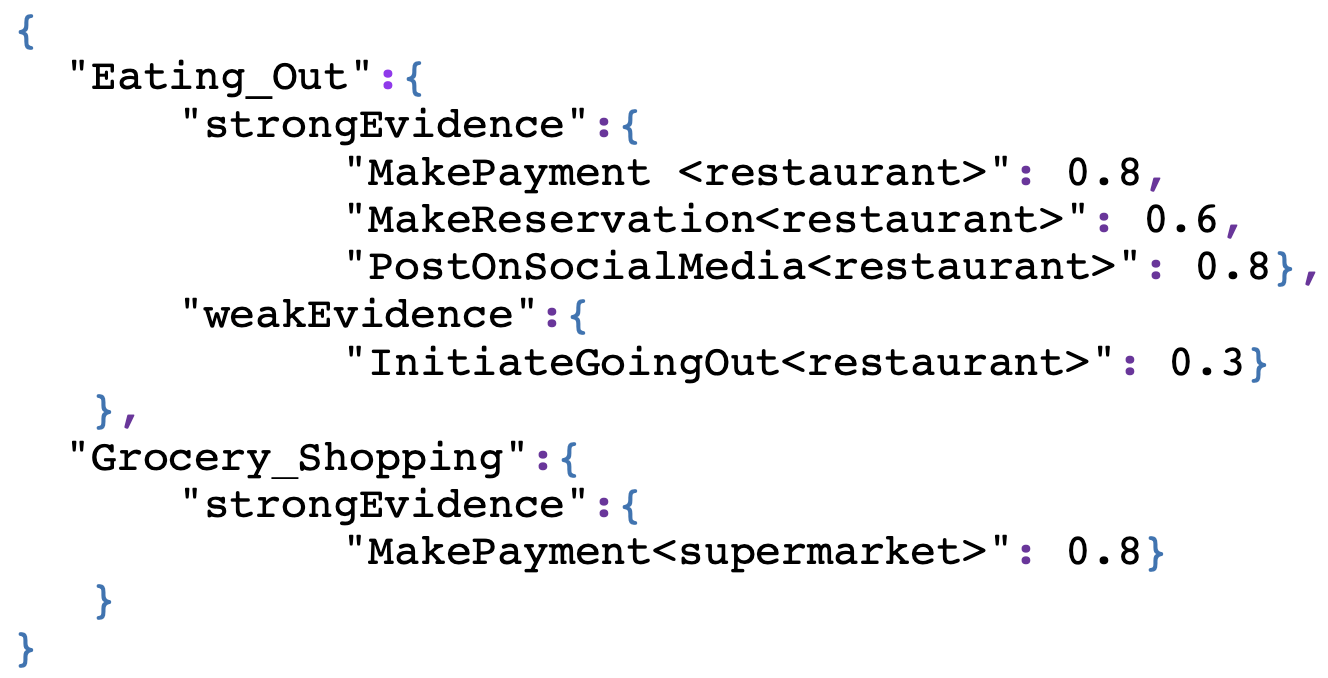}
\caption{Declarative evidence for the Eating\_Out script.}
\label{fig-evidence}
\end{figure}

The declarative description of evidence strength is illustrated in Figure~\ref{fig-evidence}. 
Strong evidence usually includes occurrence of the goal event (AttendEatingOut in this case), which may in turn have its own strong evidence (MakePayment$<$restaurant$>$). An example of weak evidence event in this case is initiateGoingOut$<$restaurant$>$.

For retrieving documents that correspond to an occurrence of some evidence, we must then identify the \textit{clues} to search for in the documents. These clues are either verbs to search for in the documents (e.g., ``eat'', ``eat out'' in an email, for identifying InitiateGoingOut $<$restaurant$>$) or specific types, attributes, and metadata that a document may have (e.g., the category in a bank card statement should be ``Restaurant'' vs ``Supermarket'' for finding evidence for the makePayment$<$restaurant$>$ vs makePayment$<$supermarket$>$).
In order to make this easily replicable for various scripts, we use
standard sources of synonyms and hyponyms like WordNet and ConceptNet5 \cite{miller1995wordnet,liu2004conceptnet}. In addition, we must consider the \wfiveh{} participants of this script/atomic action (or more specifically its FrameNet frames) in order to find additional words to search for. The words/phrases in the above generated lists are stored in a text file and are used in order to retrieve all potentially relevant documents (KEYWORDS\_FILE).

\begin{figure}[ht!]
\centering
\includegraphics[width=0.7\textwidth, height=0.5\textheight]{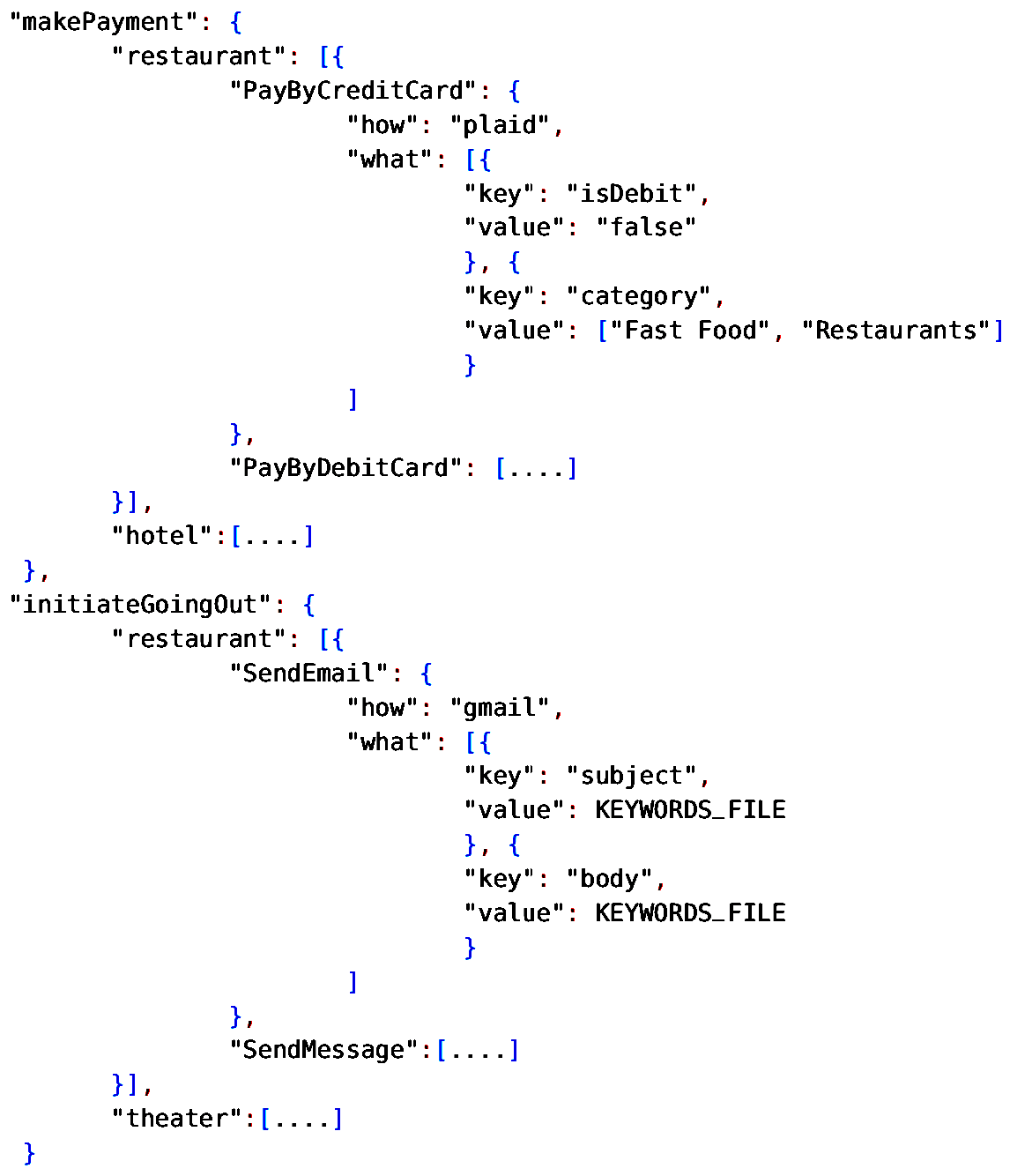}
\caption{Declarative Definition of the Clues for the makePayment and initiateGoingOut scripts.}
\label{fig-clues}
\end{figure}

The result is a list of terms to search for. For Eating\_Out, the list includes terms like ``breakfast'', ``lunch'', ``dinner'', and ``restaurant'', plus hyponyms. Figure~\ref{fig-clues} shows two examples for the clues to search for in the documents, for the MakePayment and the InitiateGoingOut sub-scripts for the restaurant script. In the former, the values to search for are based on the metadata of the document, whereas in the latter script, the KEYWORDS\_FILE corresponds to the list of terms produced with the process described above. 

We point out the ease of extending the system to recognize new scripts. For example, for makePayment$<$hotel$>$, one would just change the value ``Hotel'' for the key ``category'' of a bank transaction;  for initiateGoingOut$<$theater$>$, one would specify a KEYWORDS\_FILE starting from ``theater'' and having terms extracted from WordNet, ConceptNet, etc. The rest of the algorithm remains the same, since the queries for retrieving relevant documents are based on this declarative definition of the clues.

Finally, the set of documents D is pre-processed by: (i) explicating information (e.g., terms like ``tomorrow'', ``on Friday'', are made absolute dates) using the Natty date parser \cite{natty}; (ii) performing entity resolution for people and places (who and where dimension) using Stanford's Entity Resolution Framework \cite{serf}; (iii) grouping certain kinds of documents (e.g., related email threads, or related sequences of tweets) into a single individuals d in D; (iv) finding the places/venues that the user has visited from the geo-location coordinate history (gps) \cite{li2008mining}. 

\subsubsection{Creating initial script instances $c_d$}
Each individual document d in D instantiates the corresponding atomic action/sub-script, which leads to the creation of a candidate instance of the outer script S in a bottom-up fashion. In addition, every \wfiveh{} property is propagated from the document into the atomic actions and then into the script hierarchy above.
\tr{}
{For example, if a document of a restaurant charge denotes that it was a credit card transaction (through its metadata), the PayByCreditCard atomic action will be instantiated, along with the MakePayment sub-script, the AttendEatingOut sub-script and finally the upper-level Eating\_Out script. Then, the same document will fill the PayByCreditCard action's \wfiveh{} properties (whoPaid, whenPaid, wherePaid) and finally the outer Eating\_Out script's whereEatingOccurred, whenEatingOccurred, and one whoAttended value (the card-holder) \wfiveh{} properties.
}

\tr{}{
 Since some \PDT{}s come from events that provide strong evidence that the person was actually engaged in an instance of a specific script than others, every document d in D gets an associated initial base $score_0(d,S)$, reflecting the strength of the evidence that d manifests for being an instance of script S, and is predefined depending on the top-level episode, as shown in Figure~\ref{fig-evidence}. For example, if a credit card charge is retrieved from the strong evidence makeAPayment$<$restaurant$>$ for the Eating\_Out script, that means that the person paying at a restaurant venue most probably ate from that restaurant, giving a score for that specific document closer to 1, e.g. $score_0("bankDoc","EatingOut")=0.8$. The reason this score is not equal to 1, is because the user might have taken take out food, or someone else paid with his/her credit card. For that reason, it is really important to recognize and group other documents that account for other sub-scripts and atomic actions, in order to have multiple evidence. On the other hand, a single received email mentioning the word ``restaurant'' or ``lunch'' is extremely weak evidence, e.g. $score_0("emailDoc","EatingOut")=0.3$. However, in another type of script, where bank transactions are not important, other kinds of documents will have a greater score.

 The strength of the evidence is also affected by other factors, such as the location and number of keywords (e.g., keywords in the Subject of an email vs in the body). 
We combine compute the score of a document with multiple evidence using  Hooper's rule \cite{shafer1986combination} for combining probabilistic evidence:
 $score_t(d,S)=1 - (\prod_{i=0}^{t-1} 1 - score_i(d,S))$
 where $score_0(d,S)$ is the initial base score and
$score_{i}(d,S)$, is the score calculated from the i-th occurred evidence. Once the score for the document has been calculated, its corresponding candidate instance will get created with score equal to the score of the document for that script S, $score(S) = score_t(d,S)$. 
}

\subsubsection {Merging script instances from MergeSet} A distinctive features of our system is the presence of multiple sources of evidence for the same script instance. In order to merge them and find what they correspond to in the real episode, every script needs to have \textit{"keys"}, a rating of how well \wfiveh{} (sub)properties identify instances. For Eating\_Out, keys are whereEatingOccurred, whenEatingOccurred, and, to a lesser extent, who. The what and how properties of this script are not important because they would often lead to incorrect merging\tr{}{ e.g., two instances of eating sushi (what) need not be merged}.  When two instances of a script share the same/similar ``keys'' (some keys, such as time and place can be assessed for similarity using distance) they become candidates for merging. 
For elements of MergeSet, the \wfiveh{} property fillers are unioned, and the score for the merged instance is computed, using  Hooper's rule \cite{shafer1986combination} for combining probabilistic evidence:
 $score(S) =  1 - (\prod_{s\in S_0} 1 - score(S))$. 

Note that the above cannot be done in a single step: having established that a script instance is occurring with a certain degree of certainty, additional \PDT{}s can be gathered as part of the script instance when examining the script definition. For example, an Uber receipt might be added to the script instantiation as a result of the ``GoToPlace$<$restaurant$>$'' sub-script once an instance of Eating\_Out has been created. Note that this sub-script could not initiate on its own an episode of Eating\_Out, since Uber receipts can be part of many diffirent episodes.
 As part of future work we plan to dynamically learn the scoring function based on user's data and relevance feedback.

\section{Experiment Design}
\label{sec:settings}
To evaluate the efficacy of our approach we ran experiments on real users' data, where our goal was to find instances of them going out to eat at restaurants. We used this script example because it has several different \PDT{}s that get generated, and is similar to many other entertainment scripts such as going to a theater, going to a concert etc. Performing experiments on Personal Data is difficult due to the sensitive nature of some data, and the difficulty of getting personal data sets for research purposes. 
We used our extraction tool proposed in \cite{kalokyri2018yourdigitalself} to identify and retrieve data from current popular services and sources of \PDT{}s.

\subsection{Apparatus} 
We used an Android 7.0 app, while ensuring compatibility for later versions. The app  collected \PDT{}s from various services, stored them locally on the user's device, and  allowed the user to report back to us via prompts. The \PDT{}s collected by the app are from the following services:
 \begin{itemize}
     \item {Messaging: Messenger, Phone text messages}
     \item {Social Media: Facebook, Instagram}
     \item {Email: Gmail}
     \item {Calendar: Google Calendar}
     \item {Financial Data: Plaid API, directly downloaded .csv files from bank institutions}
     \item {Location Data: Google Maps location history}
     \item {Photos: Google Photos}
 \end{itemize}

Note that we did not have access to the user's data.
The app automatically uploaded the collected reports to a remote Firebase server via a background service. 

 \begin{figure*}[ht!]
\centering
\begin{minipage}{\textwidth}
\centering
\hidegraphics{
 \includegraphics[width=0.4\textwidth,height=4.2cm]{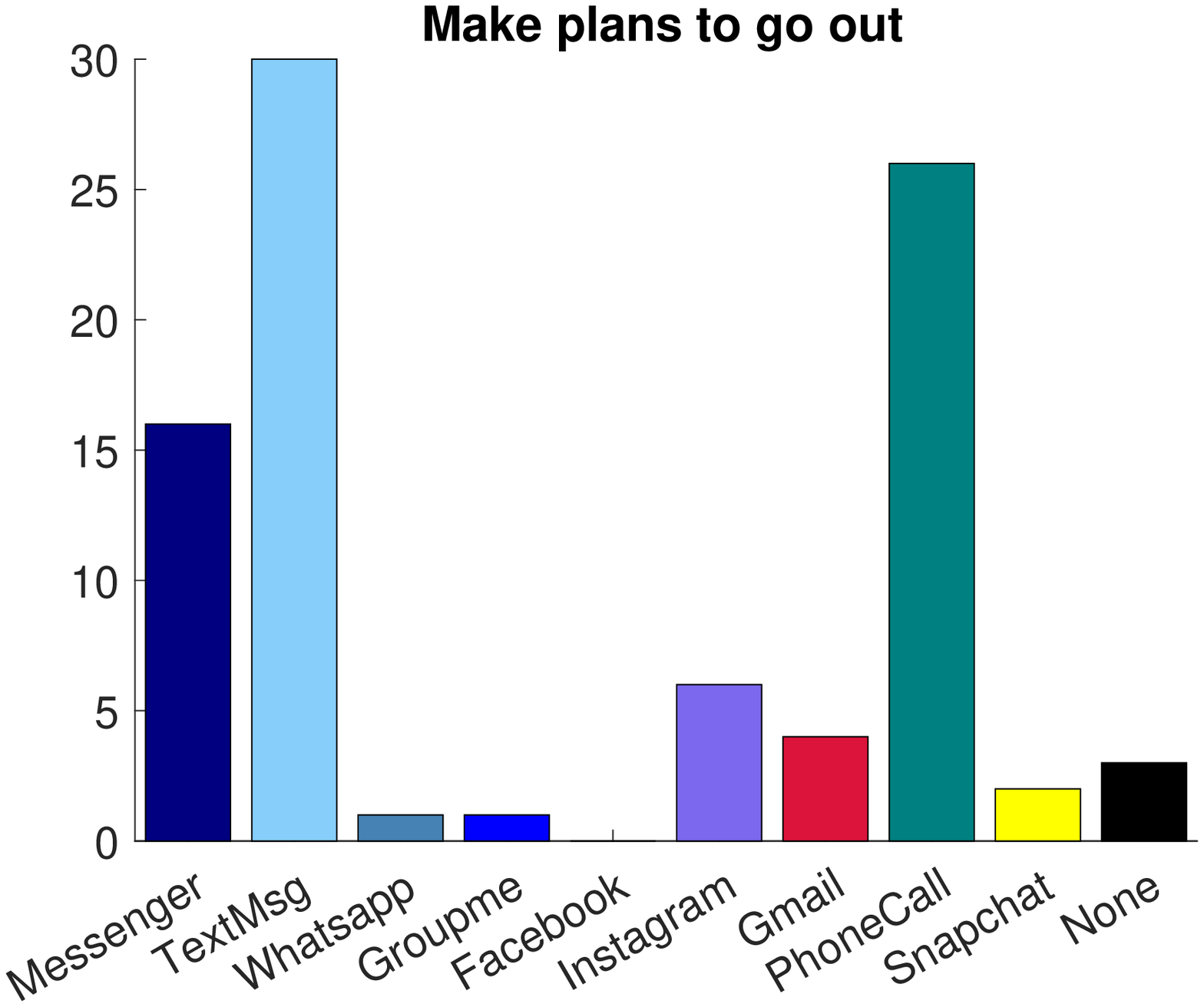}
 \includegraphics[width=0.4\textwidth,height=4.2cm]{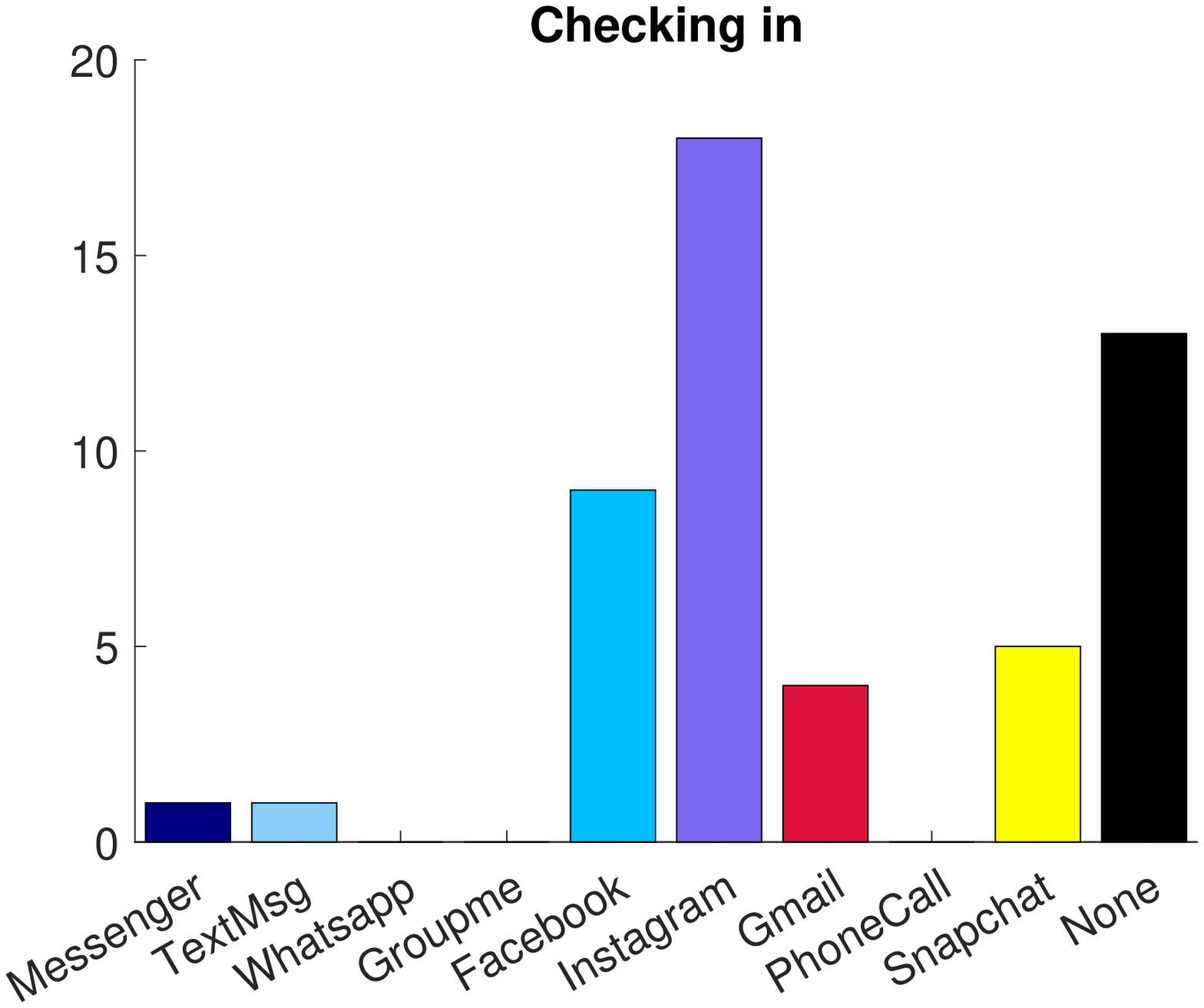}
}
\end{minipage}
\begin{minipage}{\textwidth}
\centering
\hidegraphics{
  \includegraphics[width=0.3\textwidth]{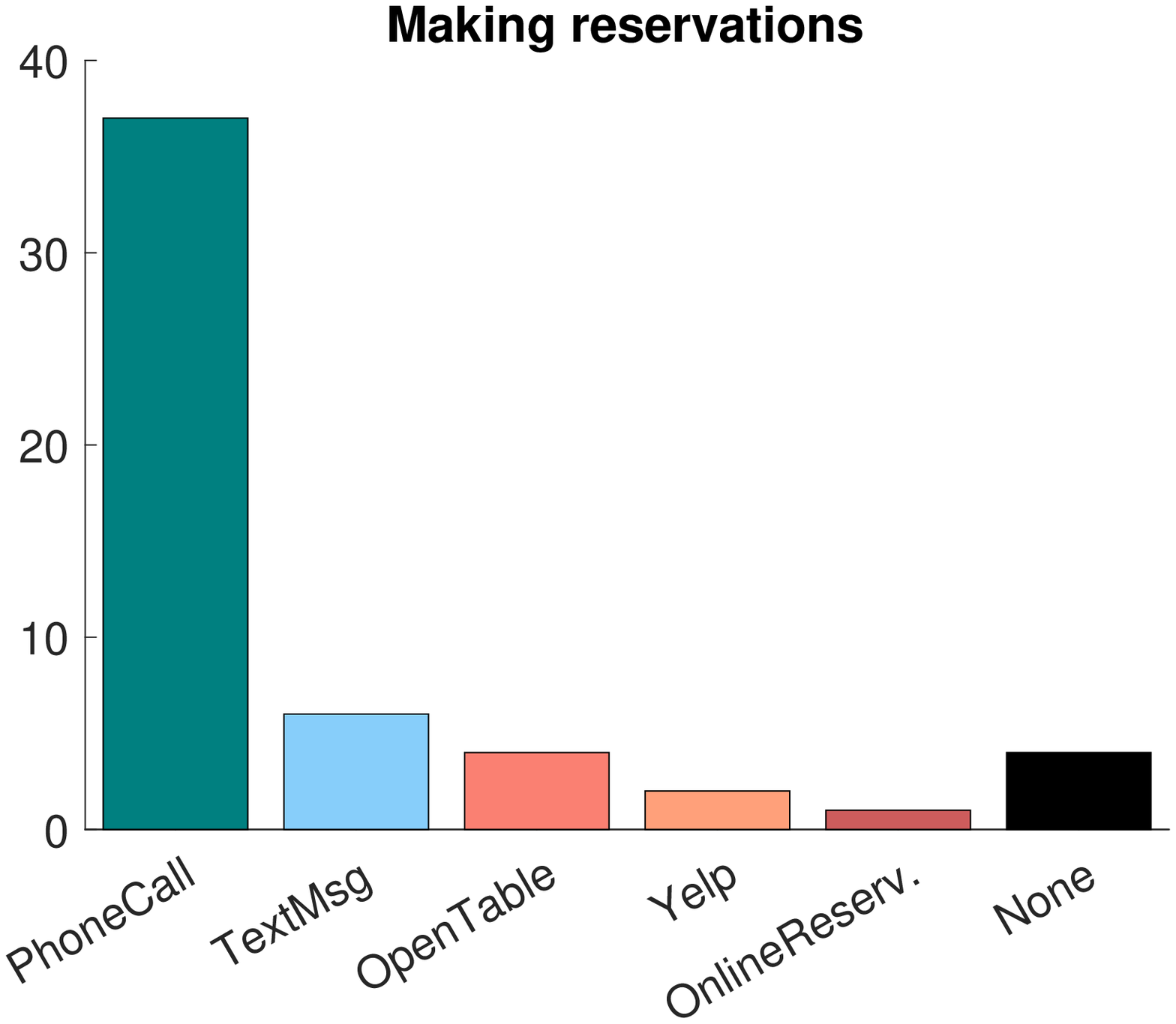}
  \includegraphics[width=0.28\textwidth]{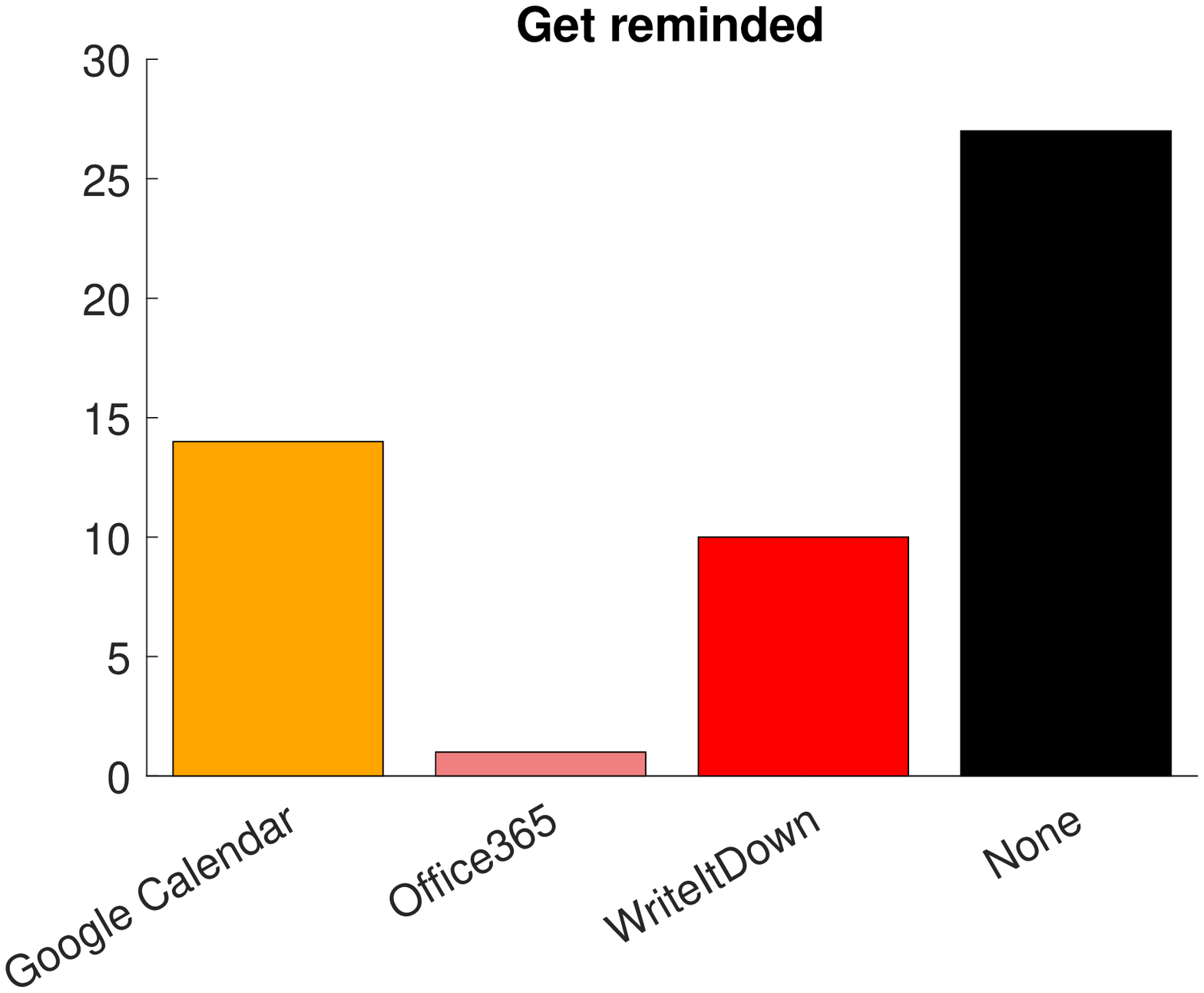}
 \includegraphics[width=0.3\textwidth]{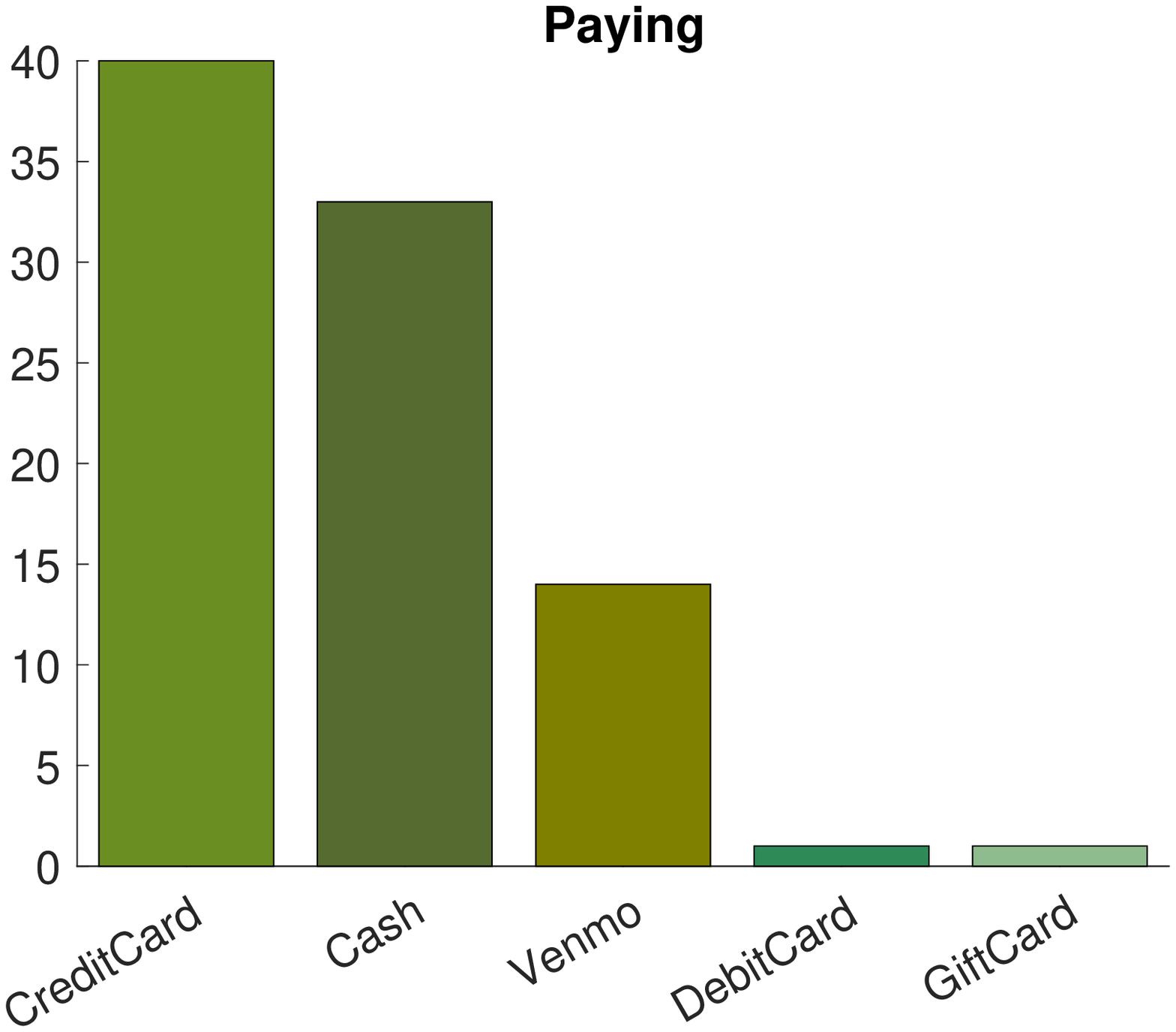}
}
\end{minipage}
\caption{Sources used by users for the Eating\_Out script.
  \emph{Top-Left:}To make plans to go out eat, \emph{Top-Right: } To let friends know that they are at a restaurant, \emph{Bottom-Left: }To make reservations at restaurants,
  \emph{Bottom-Middle: } To get reminded of restaurant outings, \emph{Bottom-Right: }To pay at restaurants}   
\label{fig:plots-sources}
\end{figure*}

\subsection{Procedure} 

We recruited participants using flyers and email lists at XXX University. We required the participants to be at least 17 years old, to be active Android users, and to communicate in English with friends and family. The participants were compensated with cash for completing the whole study. The study was approved by the authors' institution's IRB committee. 

Prospective participants took an initial survey to assess their use of potential services for Eating\_Out supported by our tool through a Likert scale (1-5).
\tr{}{
They were also asked if they were active users (as far as going out to eat). 
They were asked to reply through a Likert scale (1-5) and full text about:
\begin{itemize}
	\item{how much they think they use each of the services that the app supports}
	\item{how often they go out to eat at restaurants per month}
	\item{what services/apps they use to make plans to go out to eat, make reservations, get reminded of outings, pay and checking in at restaurants (i.e. the sub-scripts of the Eating\_Out script)} 
	\item{which services they are willing to give access to the app.}
\end{itemize}
}
We received 42 responses in total. Figure~\ref{fig:plots-sources} shows the different \PDT{} sources reported by participants for the five main sub-scripts of  Eating\_Out. A first observation is that users clearly behave differently. Most use either plain text messages or phone calls to arrange to go out to eat. 
\tr{}{
Other users use Messenger, Instagram messages or emails in order to communicate with their friends concerning this matter, and a smaller percentage uses Snapchat, Whatsapp and Groupme.
}
 On the other hand, users seem to agree in the way they make reservations (by phone) and pay at restaurants (with credit card or cash). In addition, users seem to not write down their restaurant outings in a digital form.\tr{}{
, since just 36\% of the users uses either Google Calendar or Office, and  88\% claims that they do not use any online service, with  73\% of it claiming they do not use anything to remember their restaurant outings.
}
 Finally, users seem to use many different sources for letting their friends/followers know that they are at or went to a restaurant.
\tr{}{
 The majority claims they use Instagram first and then Facebook, whereas 31\%  claims that they do not use any online service.
}
These results show that looking at several sources of \PDTs{} to identify script instances for a given user is critical, and that 
any approach to retrieve user memories must consider multiple sources of \PDTs{} to adapt to the wide variety of user behaviors.

\subsection{Experimental Setup} 
Out of the 42 recorded responses, we selected 16 participants for an in-depth study. Participants were selected based on their interest, willingness to use the app, use of services included in the app, and frequency of restaurant outings (at least 5 per month).  
 Out of the 16 participants, 9 were male and 7 female, between  ages 19 and 49, and the study was carried out with one month's data. The steps that we followed were the following:
\begin{enumerate}
    \item Before the experiment, we asked participants to carefully go over their past month's digital information, and identify the occasions they went out to eat at restaurants, including
{}{name of the restaurant --- where, date they went --- when, with whom they went --- who}. We used this information as a proxy for recall. 
\item We introduced the participants to the experiment, and we installed the app on their phone.
\item Participants were asked to give permission to the app to download and use one month's \PDT{}s of services they wanted. 
\item Participants were shown all candidate script instances of them going out to eat at restaurants, and had to indicate Yes/No for each of the instances, with further Yes/No questions as follows:
\begin{itemize}
\item Yes: evaluating the \wfiveh{} information deduced
\begin{enumerate}[label=\arabic*.,ref=\theenumi.arabic*]
	\item Who: All the identified people are correct, but there are some people missing. 
	\item Who: Some of the identified people are incorrect. 
	\item Who: All the identified people are incorrect.
	\item When: The date identified is wrong. 
	\item Where: The place name is wrong.
	\item Other. Please specify.
	\end{enumerate}
\item No: choose a reason why not. 
	\begin{enumerate}[label=\arabic*.,ref=\theenumi.arabic*]
	\item This is not a restaurant. 
	\item Take out order at that restaurant.
	\item Someone else went to that restaurant.
	\item It is a restaurant but I didn't go there to eat.
	\item Other. Please specify.
	\end{enumerate}
\end{itemize}
\item We uninstalled the app from the users' phones and destroyed the associated data.
\item We asked participants' opinions and comments at the end of the experiment.
\end{enumerate}

\section{Experimental Evaluation}
\label{sec:results}

We now detail the results of our evaluation, looking at the quality of script instance recognition.
\subsection{Experimental Setting}
We evaluate our matching algorithm and scoring function, by reporting two kinds of results: (1) how well our approach recognized script instances, and (2), how well it recognizes and abstracts When, Where and Who information from sub-scripts and atomic actions into the (outer) script instance.
For both of these cases, we report two different kinds of relevance: 
\begin{enumerate}[topsep=0pt]
\item \textbf{Binary Relevance} 

Here, a proposed instance is judged relevant if the user actually did going eat out to a restaurant, even if the \wfiveh{} information was only partly correct. Cases like take-out\tr{}{, or users going to a restaurant but not eating} were counted as false positives in this experiment. 

Each \wfiveh{} information was judged relevant only if it was exactly correct (subset or superset did not count). For example, if in the Who dimension only a subset of the people that attended was recognized, then this would be irrelevant. 

\item \textbf{Graded Relevance} 

A proposed script instance is
\begin{itemize}
\item  \textit{Exactly relevant:} when the user actually went out to eat at that  restaurant.
\item  \textit{Relevant but too broad:} when a restaurant outing is identified, but the user didn't go there to eat (e.g. went for dancing, just hanging out with a friend etc). 
\item  \textit{Relevant but too narrow:}  when a restaurant outing is identified, but the user didn't stay at the place to eat  (e.g. it was a takeout order).
\item  \textit{Partially relevant:}  when a planned outing was
correctly identified, but the user didn't end up going in the end.
\item  \textit{Not relevant:} when the identified instance is not about a restaurant.
\end{itemize}

We assume that a piece of (when, where who) \wfiveh{} information is:
\begin{itemize}
\item  \textit{Exactly relevant:} when the all identified \wfiveh{} information is correct.
\item  \textit{Relevant but too broad:} when the \wfiveh{} information contains relevant information but also includes other irrelevant information.
\item  \textit{Relevant but too narrow:}  when the \wfiveh{} information contains relevant information but is lacking some information.
\item  \textit{Not relevant:} no relevant information.
\end{itemize}
\end{enumerate}

\subsection{Metrics}
Based on both the binary and graded relevance, we report on the following metrics:
\begin{itemize}
\item \textbf{Percentage of instances retrieved:} the percentage of all Eating\_Out events identified by users which were retrieved by our scripts;  a proxy for Recall. 
\item \textbf{Mean Average Precision @ k (MAP@k):} MAP as a binary relevance assessment for the percentage of the top-k identified script instances that correspond to actual Eating\_Out events. We assumed the result counted as true positive only if the users annotate a search result as ``Exactly Relevant''. \tr{ All the other cases, are counted are false positives.}
The same holds for the \wfiveh{} information. 
\tr{}{Only when all the information identified per property is correct, the result is a true positive.}
\item \textbf{Normalized discounted cumulative gain (nDCG): } nDCG for assessing the ranked results when taking into consideration graded relevance, as described above. For our experiments, we translate the five grades of relevance as follows: Exactly relevant has a score of 5, Relevant but too broad and Relevant but too narrow have a score of 4 and 3 respectively, Partially relevant has a score of 2, and Not relevant has a score of 1. 
\end{itemize}

\begin{table*}
\begin{center}
{\small 
\begin{tabular}{|l||c|c|c|c|c|c|c|c|c|c|c|c|c|c|c|c|}
\hline
  & \#1 & \#2 & \#3 & \#4 & \#5 & \#6 & \#7 & \#8 & \#9 & \#10 & \#11 & \#12 & \#13 & \#14 & \#15 & \#16\\
\hline
\hline
Our approach  & 13 & 15 & 10 & 5 & 14 & 24 & 13 & 6 & 19 & 17 & 19 & 7  & 11 & 5 & 17 & 15\\
\hline
 User  & 13  & 14  &  11 & 5 & 7 & 14 & 11 & 6 & 20 & 14 & 15 & 5 & 11 & 5  & 16 & 14\\
\hline
\hline
\textbf{Recall}  & 1 & 1 & 0.91 & 1 & 1 & 1 & 1 & 1 & 0.95 & 1 & 1 & 1  & 1 & 1 & 1 & 1\\
\hline
\end{tabular}
}
\caption{Number of identified Eating\_out actions by users vs number of correct events our approach retrieved per user as a proxy for Recall}
\label{table:events-retrieved}
\end{center}
\vspace{-10pt}
\end{table*}

\begin{table*}
\setlength{\tabcolsep}{3pt} 
\begin{center}
{\small 
\begin{tabular}{|l||c|c|c|c|c|c|c|c|c|c|c|c|c|c|c|c|c|}
\hline
  & Sources &  \textbf{Precision}\\
\hline
\hline
User \#1 & Social Media, Calendar, Financial Data, Location Data, Google Photos & 0.87 \\
User \#2 & Social Media, Location Data, Financial Data &\textbf{0.94} \\
User \#3 & Email/Messaging, Social Media, Calendar, Financial Data & 0.66\\
User \#4 & Email/Messaging, Social Media, Calendar, Financial Data, Location Data, Google Photos & 0.66 \\
User \#5 & Email/Messaging, Social Media, Calendar, Financial Data, Location Data & 0.74\\
User \#6 & Social Media, Calendar, Financial Data, Location Data, Google Photos & 0.89 \\
User \#7 & Email/Messaging, Social Media, Calendar, Financial Data, Location Data, Google Photos & 0.76\\
User \#8 & Email/Messaging, Social Media, Calendar &\textbf{0.6} \\
User \#9 & Email/Messaging, Social Media, Calendar, Financial Data, Google Photos & 0.74 \\
User \#10 & Email/Messaging, Social Media, Calendar, Financial Data, Location Data & 0.81\\
User \#11 & Email/Messaging, Social Media, Calendar, Financial Data, Location Data, Google Photos & 0.76\\
User \#12 & Social Media, Calendar, Financial Data, Location Data, Google Photos & 0.86\\
User \#13 & Email/Messaging, Social Media, Calendar, Financial Data, Google Photos & 0.73\\
User \#14 & Email/Messaging, Social Media, Calendar, Financial Data, Location Data, Google Photos&  0.83\\
User \#15 & Email/Messaging, Social Media, Calendar, Financial Data, Google Photos & 0.77\\
User \#16 & Email/Messaging, Social Media, Calendar, Financial Data, Location Data, Google Photos & 0.79\\
\hline
\textbf{Total} & &\textbf{0.78}\\
\hline
\end{tabular}
}
\caption{Overall precision for each user}
\label{table:precision}
\end{center}
\end{table*}

\subsection{Experimental Results}

Our results allow us to make several observations.

\subsubsection{Routine experiences are hard to retrieve.}

Table~\ref{table:events-retrieved} shows the number of correct Eating\_Out instances retrieved by our approach compared with the number
identified by users from memory, and by searching their \PDT{}s (proxy for Recall). A first observation is that the results clearly indicate how hard is for users to recall their outings in the previous month, either from memory, or even when asked to go through their digital information. Our tool identified more correct instances than the users recalled in all but two cases (user 3 and 9), where the recall was 0.91 and 0.95 respectively. We anticipated this for two reasons. First,  there is evidence \cite{bradburn1987answering} that routine experiences, like going out to eat, are less likely to be encoded and harder to be retrieved, whereas unique experiences are particularly likely to be encoded and retrieved. Second, users had a hard time reviewing their digital information since they had to look in so many different services.  
Users found it very hard to search through their data properly  when using only text messages to arrange the sortie, paying by cash, or without having their GPS location activated, since most of the applications have keyword-based search. Users 5 and 6, who initially recalled only half of their outings, clearly showed this issue.

This finding supports how helpful our system can be not only for supporting human memory but also for contact tracing for epidemiological purposes. It can help users remember when they went out to eat, where and with whom. Similarly, using the grocery shopping script, the system could reveal if a user took the train/bus to go shopping (if this was instantiated from the ``GoingToPlace'' sub-script), the times they went, and to which grocery stores. 

\subsubsection{Quality of information given by different sources vary.}
Table~\ref{table:precision} shows the overall precision of the identified script instances along with the sources each user incorporated in the application. Our approach achieves a total of 78\% for all the users. User \#2 achieved the highest precision of all, since the sources they chose to include in the study contained bank transactions, google maps location history, instagram and facebook, sources that tend to be of high quality, whereas User \#8 achieves the lowest precision of all, since they included their private phone text messages without any high quality source, such as location or bank data. The reason why text documents tend to be of lower quality is because they depend on keyword matching for relevance. In addition, cases that users are discussing about going out to eat, but at the end their outing is cancelled, are hard to be recognized without advanced NLP techniques (which we are not performing in the current version of our system, e.g., we miss cases of ``cannot make it'', or, ``I will pass for today'').

The results above show that our scripts achieve good precision. However, retrieval systems typically return results in a ranked order, and users are expecting the first few results to be the most relevant.
We now look at the quality of the returned answers by evaluating the Precision@k metric. 

\subsubsection{Quality of the returned answers.}

\begin{figure}
\centering
\begin{minipage}{.5\textwidth}
  \centering
  \includegraphics[width=0.8\linewidth]{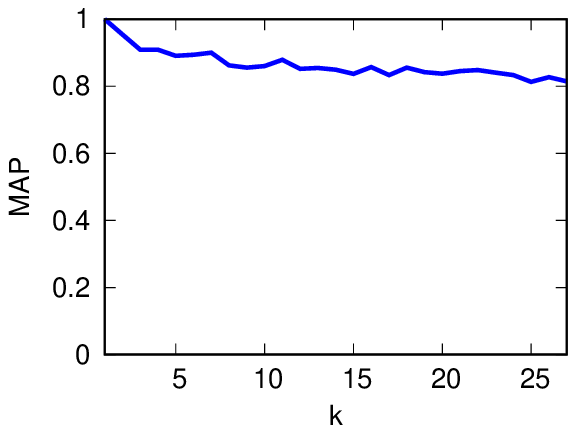}
  \captionof{figure}{Mean Average Precision@k for the \newline recognized instances for all users}
  \label{fig-MAP}
\end{minipage}%
\begin{minipage}{.5\textwidth}
  \centering
  \includegraphics[width=0.8\linewidth]{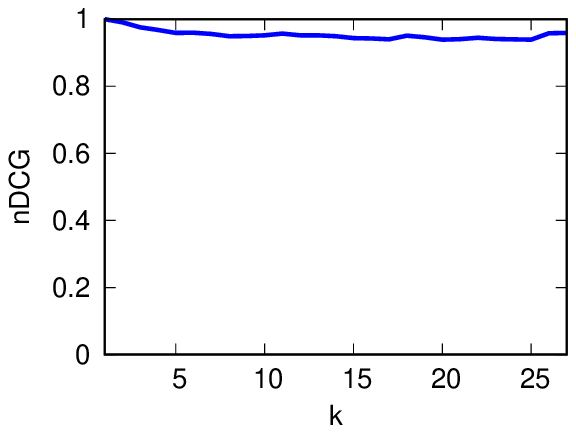}
  \captionof{figure}{Normalized Discounted Cumulative Gain @k  for the recognized instances for all users.}
  \label{fig-nDCG}
\end{minipage}
\end{figure}

\begin{figure}
\centering
\begin{minipage}{.5\textwidth}
  \centering
  \includegraphics[width=0.8\linewidth]{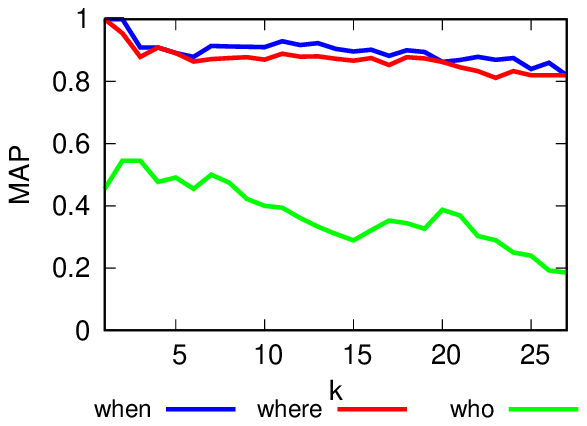}
  \captionof{figure}{Mean Average Precision @k for when, \newline where, who dimensions for all users.}
  \label{fig-w5hMAP}
\end{minipage}%
\begin{minipage}{.5\textwidth}
  \centering
  \includegraphics[width=0.8\linewidth]{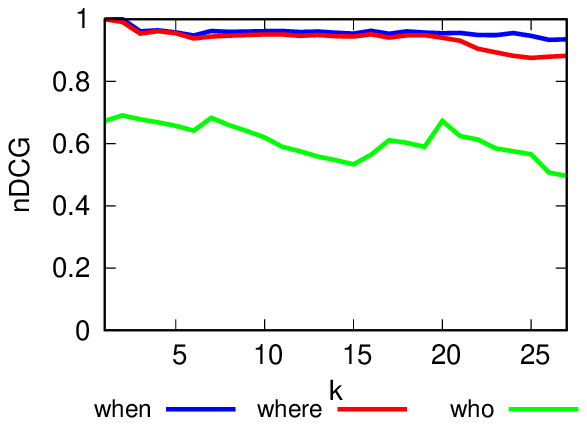}
  \captionof{figure}{Normalized Discounted Cumulative Gain @k for when, where, who dimensions for all users.}
  \label{fig-w5h_nDCG}
\end{minipage}
\end{figure}

Figure~\ref{fig-MAP} shows the Mean Average Precision@k for all the identified script instances for all the users. As shown, our approach achieves a really good precision even for low values of k. The reason for that, is that our approach does include many different kinds of sources and is able to account for all the different kinds of user behavior. In our previous study we showed how important it is to incorporate multiple sources of \PDT{}s in order to recognize user's actions  \cite{usexploreDB}. In addition, the users seem to use different services for carrying out different actions of a particular script instance, as noted in Figure~\ref{fig:plots-sources}. 

Figure~\ref{fig-nDCG} shows the normalized discounted cumulative gain (nDCG) for the ranked results when taking into consideration the graded relevance. The nDCG was computed by normalizing the DCG@k with the ideal DCG value or IDCG@k. Hence, we computed the IDCG at each level k and then computed the average nDCG across the 16 results. It is clear that our ranking quality is high, and our approach is able to recognize and distinguish highly relevant \PDT{}s in favor of irrelevant \PDT{}s.

We then report the same metrics (MAP, MAP@k and nDCG@k) on the when, where and who dimensions.

\begin{table}
\begin{center}
{\small 
\begin{tabular}{|l||c|c|c|}
\hline
 & when & where & who \\
\hline
\hline
MAP& 0.85 & 0.81 & 0.21 \\
\hline
\end{tabular}
}
\caption{MAP for when, where, who dimensions for all users}
\label{table:w5h_MAP}
\end{center}
\vspace{-10pt}
\end{table} 

\subsubsection{The who dimension is harder to be retrieved.}
Table~\ref{table:w5h_MAP} shows the MAP for the three dimensions for all the users. A first observation is that the when and where dimensions are easier to extract than the who dimension due to the metadata that the \PDT{}s have and due to the fact that if there is a payment or a gps location for an outing these two dimensions are easier to get extracted. On the other hand, the who dimension is more difficult to get extracted for the following reason. Most of the participants' restaurant outings as shown in Figure~\ref{fig:plots-sources}, were arranged either by phone or by text messages. Our system does not capture voice data, so it was anticipated that we will miss some information about the phone arranged outings. In addition, the participants mentioned many times that many of their outings were organized on the fly, by talking to each other in person, while in work, or while being together with friends. Our evaluation also penalizes correct but incomplete who dimensions by counting them as false positives. 

Figure~\ref{fig-w5hMAP} shows the MAP@k for the three dimensions for all the users. As shown, our approach achieves a really good precision for the when and where dimension for all the values of k. On the other hand, the precision for the who dimension drops as k increases. This happens due to the fact that for low values of k, the score of the instances is low, which means there are not many \PDT{}s to account for these instances. In that case, as previously mentioned, the who dimension is the hardest dimension to be retrieved, because either there is no information recorded or our approach either lacks or recognizes more people in an outing. This is  actually demonstrated in Figure~\ref{fig-w5h_nDCG}, which shows the nDCG for the ranked results when taking into consideration the graded relevance.  We can now observe how much better the accuracy is for the who dimension getting a gain of 0.7 as the highest and 0.5 as the lowest value. This means that our approach is able to recognize some people in the outing, but it's hard to recognize them all correctly. 
In addition, we observe that the when dimension is getting extracted with a better accuracy than the where dimension. This is mainly because the when dimension can be extracted by all the \PDT{}s, whereas the where dimension can be missing from text messages, emails or photos. 
\subsubsection{Our system was rated positively by the participants.} In general, users were satisfied with the efficiency of use of the application and the functionality it offers. The majority was really exciting about using the app. They mentioned how "cool" it is to be able to see their data organized, and how easy it is to navigate to their data, since it is linking to the original piece of information. In addition, some were surprised when they found out some outings that they had totally forgotten. They pointed out that this happens to them frequently, and that although they do remember having some pieces of information somewhere in some service, they don't know where to search, or what to search. Finally, some mentioned that they would prefer to be able to search through our app, rather than browsing their instances, as well as they wished that some other services were supported by the app (i.e. Snapchat, WhatsApp).

To sum up, our results showed that our approach is able to integrate and connect different data traces into script instantiations and that any approach for recognizing user activities should include many different kinds of sources in order to be able to account for all the different kinds of user behaviors. In addition, we showed evidence that our approach does augment the users memories of their past actions, even if these actions were relatively recent (within the past month).  Finally, our work can be leveraged to help users organize and remember information  for their personal use, but it could also assist health officials to perform contact and location tracing.

\section{Related Work}
\label{sec:related}

We review the many areas of related work, some of which have already been mentioned briefly in the previous sections.

\textbf{ Activities of Daily Living, Pervasive Computing, LifeLogging and Memory Tools.} 
The use of aids to help people with memory deficits is thought to be one of the most effective ways to aid rehabilitation (see \cite{kapurexternal} for a review).
 Most external memory aids focus on improving prospective memory; they help people to remember to keep appointments, take medication, etc. Available devices include calendars, alarms,  Post-It notes, as well as more sophisticated systems, like Amazon Alexa, Siri, and Google Now \cite{hoy2018alexa, thakur2016personalization}. 
In contrast, there are few memory aids designed to improve the ability to remember past experiences. Perhaps the two most obvious examples are cameras and diaries. Sensecam \cite{hodges2006sensecam} is a tool closely related to ours, used for the recall of everyday events, by passively recording images relating to everyday activity in order to trigger autobiographical memory in people with memory issues. Other tools include the MemoClip , the CybreMinder, and Memory Glasses \cite{memoclip, devaulmemory, deycybreminder}. Our work is distinguished from most of the above efforts by the fact that we use the vast amount of existing digital traces already being produced, rather than capturing new data.

In addition, the area of life-logging is quite similar, and is
surveyed in \cite{gurrin2014lifelogging, van2012introduction}. Bell has pioneered the field of life-logging with the project MyLifeBits \cite{mylifebits} for which he has digitally captured all aspects of his life. 
A particularly relevant paper is \cite{ADLasbru}, which uses the plan description
language Asbru \cite{asbru} (designed to describe
medical protocols as skeletal plans) to recognize activities
of daily leaving from kitchen sensors. 
Our approach is distinguished from the above efforts
by the facts that: (i) there is massive concurrent execution of different
 scripts; (ii)  many PDTs are not part of
any script execution; (iii) the execution of some script steps is not  manifested in PDTs; (iv) script instances are often exceptional
variants of  the prototype.

\textbf{PIM.} The general research area of Personal Information Management (PIM) began in the 1980s to help users better store, integrate and query large collections of varied digital data. Researchers have suggested PIM interfaces for web activities \cite{stuff, kaptelinin2003umea, murakami2012system}, email \cite{ayodele2012machine, balter1997strategies, bellotti2002innovation, snyder2013cloudsweeper, whittaker2006email, whittaker1996email}, and local files \cite{barreau1995finding, barreau1995context}. 
The central focus of such systems is the identification of relevant objects in the user's information space, and establishing their inter-relationships. Often this is based on a domain or personal ontology. In contrast with this static view of information, we focus on a dynamic approach to the integration of \PDT{}s, by providing a narrative to make connections between them.

\textbf{Processes and Plans.}
Since scripts are plans, and we want to recognize plan instances
from \PDT{}s, the extensive literature on plan
recognition, such as \cite{geibGoldman,geib.yappr}, is obviously relevant. One important difference is that these approaches start from a description of a domain
in terms of planning operators, while scripts are pre-compiled
stereotypical plans.
Also closely related is the areas of activity
recognition, which often considers the problem of recognizing
lower-level actions, of which plans are composed, especially
when these are signaled by sensors. 
Our needs include the ability to recognize multiple, concurrent, and interleaved script instances and components. 
\tr{}{
(e.g., ``working at the office
today'', ``planning to go to a wedding this weekend'', ``planning to have dinner guests tonight''). Also, we observe ``\PDT{}s'' rather than primitive low-level actions, or their
signature via sensors.
}
 Most importantly,  our situation is distinguished by the fact that  most of the \PDT{}s we encounter do not signal any script, and a very large fraction of steps in any
particular instantiation of a script leave no trace (``missing actions'').

\vspace{-2pt}
\section{Conclusion}
We have presented a novel script-based approach to integrate and connect heterogeneous collections of \PDT{}s into coherent episodes of user activities, which extract relevant summary information. Our approach can help users explore their events in an integrated way by creating a personal knowledge base they can access and search in the future.
Experiments on real users' personal data for the script of going out to eat 
showed that our approach augments people's memory for past actions, which can subsequently help them to stimulate their memory. Simple variants of Eating\_Out, such as Going\_To\_Theater and other forms of entertainment, would cover many more cases. This can be particularly useful in a variety of situations such as people with memory deficits, as well as contact and location tracing.
In addition to its applications to personal data management and memory augmentation, our work provides opportunities for behavioral researchers to study user behavior patterns. For example, we believe that the combination of \PDTs{}, AI techniques, and self-reported surveys can yield new insights into mental health assessment.

\bibliographystyle{ACM-Reference-Format}
\bibliography{bibliography}


\begin{thebibliography}{52}


\ifx \showCODEN    \undefined \def \showCODEN     #1{\unskip}     \fi
\ifx \showDOI      \undefined \def \showDOI       #1{#1}\fi
\ifx \showISBNx    \undefined \def \showISBNx     #1{\unskip}     \fi
\ifx \showISBNxiii \undefined \def \showISBNxiii  #1{\unskip}     \fi
\ifx \showISSN     \undefined \def \showISSN      #1{\unskip}     \fi
\ifx \showLCCN     \undefined \def \showLCCN      #1{\unskip}     \fi
\ifx \shownote     \undefined \def \shownote      #1{#1}          \fi
\ifx \showarticletitle \undefined \def \showarticletitle #1{#1}   \fi
\ifx \showURL      \undefined \def \showURL       {\relax}        \fi
\providecommand\bibfield[2]{#2}
\providecommand\bibinfo[2]{#2}
\providecommand\natexlab[1]{#1}
\providecommand\showeprint[2][]{arXiv:#2}

\bibitem[\protect\citeauthoryear{??}{ser}{[n.d.]}]%
        {serf}
 \bibinfo{year}{[n.d.]}\natexlab{}.
\newblock \bibinfo{title}{Stanford Entity Resolution Framework}.
\newblock \bibinfo{howpublished}{\url{http://infolab.stanford.edu/serf/}}.
\newblock
\newblock
\shownote{Accessed: 2019.}


\bibitem[\protect\citeauthoryear{??}{Goo}{2020}]%
        {GoogleApple}
 \bibinfo{year}{2020}\natexlab{}.
\newblock \showarticletitle{Google Inc. Apple and Google partner on COVID-19
  contact tracing technology}.
\newblock  (\bibinfo{year}{2020}).
\newblock
\urldef\tempurl%
\url{https://www.blog.google/inside-google/company-announcements/apple-and-google-partner-covid-19-contact-tracing-technology/}
\showURL{%
\tempurl}


\bibitem[\protect\citeauthoryear{Ayodele, Akmayeva, and Shoniregun}{Ayodele
  et~al\mbox{.}}{2012}]%
        {ayodele2012machine}
\bibfield{author}{\bibinfo{person}{Taiwo Ayodele}, \bibinfo{person}{Galyna
  Akmayeva}, {and} \bibinfo{person}{Charles~A Shoniregun}.}
  \bibinfo{year}{2012}\natexlab{}.
\newblock \showarticletitle{Machine learning approach towards email
  management}. In \bibinfo{booktitle}{\emph{World Congress on Internet Security
  (WorldCIS-2012)}}. IEEE, \bibinfo{pages}{106--109}.
\newblock


\bibitem[\protect\citeauthoryear{Balter}{Balter}{1997}]%
        {balter1997strategies}
\bibfield{author}{\bibinfo{person}{Olle Balter}.}
  \bibinfo{year}{1997}\natexlab{}.
\newblock \showarticletitle{Strategies for organising email messages}.
\newblock \bibinfo{journal}{\emph{HCI 1997}} (\bibinfo{year}{1997}),
  \bibinfo{pages}{21--38}.
\newblock


\bibitem[\protect\citeauthoryear{Barreau and Nardi}{Barreau and Nardi}{1995}]%
        {barreau1995finding}
\bibfield{author}{\bibinfo{person}{Deborah Barreau} {and}
  \bibinfo{person}{Bonnie~A Nardi}.} \bibinfo{year}{1995}\natexlab{}.
\newblock \showarticletitle{Finding and reminding: file organization from the
  desktop}.
\newblock \bibinfo{journal}{\emph{ACM SigChi Bulletin}} \bibinfo{volume}{27},
  \bibinfo{number}{3} (\bibinfo{year}{1995}), \bibinfo{pages}{39--43}.
\newblock


\bibitem[\protect\citeauthoryear{Barreau}{Barreau}{1995}]%
        {barreau1995context}
\bibfield{author}{\bibinfo{person}{Deborah~K Barreau}.}
  \bibinfo{year}{1995}\natexlab{}.
\newblock \showarticletitle{Context as a factor in personal information
  management systems}.
\newblock \bibinfo{journal}{\emph{Journal of the American Society for
  Information Science}} \bibinfo{volume}{46}, \bibinfo{number}{5}
  (\bibinfo{year}{1995}), \bibinfo{pages}{327--339}.
\newblock


\bibitem[\protect\citeauthoryear{Beigl}{Beigl}{2000}]%
        {memoclip}
\bibfield{author}{\bibinfo{person}{Michael Beigl}.}
  \bibinfo{year}{2000}\natexlab{}.
\newblock \showarticletitle{MemoClip: A location-based remembrance appliance}.
\newblock \bibinfo{journal}{\emph{Personal Technologies}} \bibinfo{volume}{4},
  \bibinfo{number}{4} (\bibinfo{year}{2000}), \bibinfo{pages}{230--233}.
\newblock


\bibitem[\protect\citeauthoryear{Bellotti, Ducheneaut, Howard, Smith, and
  Neuwirth}{Bellotti et~al\mbox{.}}{2002}]%
        {bellotti2002innovation}
\bibfield{author}{\bibinfo{person}{Victoria Bellotti}, \bibinfo{person}{Nicolas
  Ducheneaut}, \bibinfo{person}{Mark Howard}, \bibinfo{person}{Ian Smith},
  {and} \bibinfo{person}{Christine Neuwirth}.} \bibinfo{year}{2002}\natexlab{}.
\newblock \showarticletitle{Innovation in extremis: evolving an application for
  the critical work of email and information management}. In
  \bibinfo{booktitle}{\emph{Proceedings of the 4th conference on Designing
  interactive systems: processes, practices, methods, and techniques}}.
  \bibinfo{pages}{181--192}.
\newblock


\bibitem[\protect\citeauthoryear{Blanchette}{Blanchette}{2009}]%
        {blanchette}
\bibfield{author}{\bibinfo{person}{Jean-Fran{\c{c}}ois Blanchette}.}
  \bibinfo{year}{2009}\natexlab{}.
\newblock \showarticletitle{Total Recall: How the E-Memory Revolution will
  Change Everything. Gordon Bell \& Jim Gemmel}.
\newblock  (\bibinfo{year}{2009}).
\newblock


\bibitem[\protect\citeauthoryear{Bradburn, Rips, and Shevell}{Bradburn
  et~al\mbox{.}}{1987}]%
        {bradburn1987answering}
\bibfield{author}{\bibinfo{person}{Norman~M Bradburn}, \bibinfo{person}{Lance~J
  Rips}, {and} \bibinfo{person}{Steven~K Shevell}.}
  \bibinfo{year}{1987}\natexlab{}.
\newblock \showarticletitle{Answering autobiographical questions: The impact of
  memory and inference on surveys}.
\newblock \bibinfo{journal}{\emph{Science}} \bibinfo{volume}{236},
  \bibinfo{number}{4798} (\bibinfo{year}{1987}), \bibinfo{pages}{157--161}.
\newblock


\bibitem[\protect\citeauthoryear{Bush}{Bush}{1945}]%
        {memex}
\bibfield{author}{\bibinfo{person}{Vannevar Bush}.}
  \bibinfo{year}{1945}\natexlab{}.
\newblock \showarticletitle{As We May Think}.
\newblock \bibinfo{journal}{\emph{The Atlantic Monthly}} \bibinfo{number}{1}
  (\bibinfo{year}{1945}).
\newblock


\bibitem[\protect\citeauthoryear{Conway and Rubin}{Conway and Rubin}{1993}]%
        {conway1993structure}
\bibfield{author}{\bibinfo{person}{Martin~A Conway} {and}
  \bibinfo{person}{David~C Rubin}.} \bibinfo{year}{1993}\natexlab{}.
\newblock \showarticletitle{The structure of autobiographical memory}.
\newblock \bibinfo{journal}{\emph{Theories of memory}}  \bibinfo{volume}{103}
  (\bibinfo{year}{1993}), \bibinfo{pages}{137}.
\newblock


\bibitem[\protect\citeauthoryear{Czerwinski, Gage, Gemmell, Marshall,
  P{\'e}rez-Qui{\~n}ones, Skeels, and Catarci}{Czerwinski
  et~al\mbox{.}}{2006}]%
        {czerwinski2006digital}
\bibfield{author}{\bibinfo{person}{Mary Czerwinski}, \bibinfo{person}{Douglas~W
  Gage}, \bibinfo{person}{Jim Gemmell}, \bibinfo{person}{Catherine~C Marshall},
  \bibinfo{person}{Manuel~A P{\'e}rez-Qui{\~n}ones},
  \bibinfo{person}{Meredith~M Skeels}, {and} \bibinfo{person}{Tiziana
  Catarci}.} \bibinfo{year}{2006}\natexlab{}.
\newblock \showarticletitle{Digital memories in an era of ubiquitous computing
  and abundant storage}.
\newblock \bibinfo{journal}{\emph{Commun. ACM}}  \bibinfo{volume}{49}
  (\bibinfo{year}{2006}), \bibinfo{pages}{44--50}.
\newblock


\bibitem[\protect\citeauthoryear{DeVaul, Pentland, and Corey}{DeVaul
  et~al\mbox{.}}{2003}]%
        {devaulmemory}
\bibfield{author}{\bibinfo{person}{Richard~W DeVaul}, \bibinfo{person}{Alex
  Pentland}, {and} \bibinfo{person}{Vicka~R Corey}.}
  \bibinfo{year}{2003}\natexlab{}.
\newblock \showarticletitle{The memory glasses: subliminal vs. overt memory
  support with imperfect information}. In \bibinfo{booktitle}{\emph{Seventh
  IEEE International Symposium on Wearable Computers, 2003. Proceedings.}}
  Citeseer, \bibinfo{pages}{146--153}.
\newblock


\bibitem[\protect\citeauthoryear{Dey and Abowd}{Dey and Abowd}{2000}]%
        {deycybreminder}
\bibfield{author}{\bibinfo{person}{Anind~K Dey} {and}
  \bibinfo{person}{Gregory~D Abowd}.} \bibinfo{year}{2000}\natexlab{}.
\newblock \showarticletitle{Cybreminder: A context-aware system for supporting
  reminders}. In \bibinfo{booktitle}{\emph{International Symposium on Handheld
  and Ubiquitous Computing}}. Springer, \bibinfo{pages}{172--186}.
\newblock


\bibitem[\protect\citeauthoryear{Dumais, Cutrell, Cadiz, Jancke, Sarin, and
  Robbins}{Dumais et~al\mbox{.}}{2003}]%
        {stuff}
\bibfield{author}{\bibinfo{person}{Susan Dumais}, \bibinfo{person}{Edward
  Cutrell}, \bibinfo{person}{Jonathan~J. Cadiz}, \bibinfo{person}{Gavin
  Jancke}, \bibinfo{person}{Raman Sarin}, {and} \bibinfo{person}{Daniel~C.
  Robbins}.} \bibinfo{year}{2003}\natexlab{}.
\newblock \showarticletitle{Stuff I've Seen: A System for Personal Information
  Retrieval and Re-Use}. In \bibinfo{booktitle}{\emph{Proceedings of the 26th
  International ACM SIGIR Conference (SIGIR'03)}}.
\newblock


\bibitem[\protect\citeauthoryear{Eames and Keeling}{Eames and Keeling}{2003}]%
        {eames2003contact}
\bibfield{author}{\bibinfo{person}{Ken~TD Eames} {and} \bibinfo{person}{Matt~J
  Keeling}.} \bibinfo{year}{2003}\natexlab{}.
\newblock \showarticletitle{Contact tracing and disease control}.
\newblock \bibinfo{journal}{\emph{Proceedings of the Royal Society of London.
  Series B: Biological Sciences}} \bibinfo{volume}{270}, \bibinfo{number}{1533}
  (\bibinfo{year}{2003}), \bibinfo{pages}{2565--2571}.
\newblock


\bibitem[\protect\citeauthoryear{Geib and Goldman}{Geib and Goldman}{2009}]%
        {geibGoldman}
\bibfield{author}{\bibinfo{person}{Christopher~W Geib} {and}
  \bibinfo{person}{Robert~P Goldman}.} \bibinfo{year}{2009}\natexlab{}.
\newblock \showarticletitle{A probabilistic plan recognition algorithm based on
  plan tree grammars}.
\newblock \bibinfo{journal}{\emph{Artificial Intelligence}}
  \bibinfo{volume}{173}, \bibinfo{number}{11} (\bibinfo{year}{2009}),
  \bibinfo{pages}{1101--1132}.
\newblock


\bibitem[\protect\citeauthoryear{Geib, Maraist, and Goldman}{Geib
  et~al\mbox{.}}{2008}]%
        {geib.yappr}
\bibfield{author}{\bibinfo{person}{Christopher~W Geib}, \bibinfo{person}{John
  Maraist}, {and} \bibinfo{person}{Robert~P Goldman}.}
  \bibinfo{year}{2008}\natexlab{}.
\newblock \showarticletitle{A New Probabilistic Plan Recognition Algorithm
  Based on String Rewriting}. In \bibinfo{booktitle}{\emph{ICAPS}}.
  \bibinfo{pages}{91--98}.
\newblock


\bibitem[\protect\citeauthoryear{Gemmell, Bell, and Lueder}{Gemmell
  et~al\mbox{.}}{2006}]%
        {mylifebits}
\bibfield{author}{\bibinfo{person}{Jim Gemmell}, \bibinfo{person}{Gordon Bell},
  {and} \bibinfo{person}{Roger Lueder}.} \bibinfo{year}{2006}\natexlab{}.
\newblock \showarticletitle{MyLifeBits: a personal database for everything}.
\newblock \bibinfo{journal}{\emph{Commun. ACM}} \bibinfo{volume}{49},
  \bibinfo{number}{1} (\bibinfo{year}{2006}), \bibinfo{pages}{88--95}.
\newblock


\bibitem[\protect\citeauthoryear{Gurrin, Smeaton, Doherty,
  et~al\mbox{.}}{Gurrin et~al\mbox{.}}{2014}]%
        {gurrin2014lifelogging}
\bibfield{author}{\bibinfo{person}{Cathal Gurrin}, \bibinfo{person}{Alan~F
  Smeaton}, \bibinfo{person}{Aiden~R Doherty}, {et~al\mbox{.}}}
  \bibinfo{year}{2014}\natexlab{}.
\newblock \showarticletitle{Lifelogging: Personal big data}.
\newblock \bibinfo{journal}{\emph{Foundations and Trends{\textregistered} in
  information retrieval}} \bibinfo{volume}{8}, \bibinfo{number}{1}
  (\bibinfo{year}{2014}), \bibinfo{pages}{1--125}.
\newblock


\bibitem[\protect\citeauthoryear{Hodges, Williams, Berry, Izadi, Srinivasan,
  Butler, Smyth, Kapur, and Wood}{Hodges et~al\mbox{.}}{2006}]%
        {hodges2006sensecam}
\bibfield{author}{\bibinfo{person}{Steve Hodges}, \bibinfo{person}{Lyndsay
  Williams}, \bibinfo{person}{Emma Berry}, \bibinfo{person}{Shahram Izadi},
  \bibinfo{person}{James Srinivasan}, \bibinfo{person}{Alex Butler},
  \bibinfo{person}{Gavin Smyth}, \bibinfo{person}{Narinder Kapur}, {and}
  \bibinfo{person}{Ken Wood}.} \bibinfo{year}{2006}\natexlab{}.
\newblock \showarticletitle{SenseCam: A Retrospective Memory Aid}.
\newblock  (\bibinfo{year}{2006}).
\newblock


\bibitem[\protect\citeauthoryear{Hoy}{Hoy}{2018}]%
        {hoy2018alexa}
\bibfield{author}{\bibinfo{person}{Matthew~B Hoy}.}
  \bibinfo{year}{2018}\natexlab{}.
\newblock \showarticletitle{Alexa, Siri, Cortana, and more: an introduction to
  voice assistants}.
\newblock \bibinfo{journal}{\emph{Medical reference services quarterly}}
  \bibinfo{volume}{37}, \bibinfo{number}{1} (\bibinfo{year}{2018}),
  \bibinfo{pages}{81--88}.
\newblock


\bibitem[\protect\citeauthoryear{Huerta and Tsimring}{Huerta and
  Tsimring}{2002}]%
        {huerta2002contact}
\bibfield{author}{\bibinfo{person}{Ramon Huerta} {and} \bibinfo{person}{Lev~S
  Tsimring}.} \bibinfo{year}{2002}\natexlab{}.
\newblock \showarticletitle{Contact tracing and epidemics control in social
  networks}.
\newblock \bibinfo{journal}{\emph{Physical Review E}} \bibinfo{volume}{66},
  \bibinfo{number}{5} (\bibinfo{year}{2002}), \bibinfo{pages}{056115}.
\newblock


\bibitem[\protect\citeauthoryear{Jones}{Jones}{2007}]%
        {jones2007personal}
\bibfield{author}{\bibinfo{person}{William Jones}.}
  \bibinfo{year}{2007}\natexlab{}.
\newblock \showarticletitle{Personal information management}.
\newblock \bibinfo{journal}{\emph{Annual review of information science and
  technology}}  \bibinfo{volume}{41} (\bibinfo{year}{2007}),
  \bibinfo{pages}{453--504}.
\newblock


\bibitem[\protect\citeauthoryear{Kalokyri, Borgida, and Marian}{Kalokyri
  et~al\mbox{.}}{2018}]%
        {kalokyri2018yourdigitalself}
\bibfield{author}{\bibinfo{person}{Varvara Kalokyri},
  \bibinfo{person}{Alexander Borgida}, {and} \bibinfo{person}{Am{\'e}lie
  Marian}.} \bibinfo{year}{2018}\natexlab{}.
\newblock \showarticletitle{YourdigitalSelf: A personal digital trace
  integration tool}. In \bibinfo{booktitle}{\emph{Proceedings of the 27th ACM
  International Conference on Information and Knowledge Management}}.
  \bibinfo{pages}{1963--1966}.
\newblock


\bibitem[\protect\citeauthoryear{Kalokyri, Borgida, Marian, and
  Vianna}{Kalokyri et~al\mbox{.}}{2017a}]%
        {usexploreDB}
\bibfield{author}{\bibinfo{person}{Varvara Kalokyri},
  \bibinfo{person}{Alexander Borgida}, \bibinfo{person}{Am{\'{e}}lie Marian},
  {and} \bibinfo{person}{Daniela Vianna}.} \bibinfo{year}{2017}\natexlab{a}.
\newblock \showarticletitle{Integration and Exploration of Connected Personal
  Digital Traces}. In \bibinfo{booktitle}{\emph{Proc. ExploreDB'17}}.
  \bibinfo{pages}{3:1--3:6}.
\newblock


\bibitem[\protect\citeauthoryear{Kalokyri, Borgida, Marian, and
  Vianna}{Kalokyri et~al\mbox{.}}{2017b}]%
        {usOdbase}
\bibfield{author}{\bibinfo{person}{Varvara Kalokyri},
  \bibinfo{person}{Alexander Borgida}, \bibinfo{person}{Am{\'{e}}lie Marian},
  {and} \bibinfo{person}{Daniela Vianna}.} \bibinfo{year}{2017}\natexlab{b}.
\newblock \showarticletitle{Semantic Modeling and Inference with Episodic
  Organization for Managing Personal Digital Traces - (Short Paper)}. In
  \bibinfo{booktitle}{\emph{Proc. {ODBASE} 2017}}. \bibinfo{pages}{273--280}.
\newblock


\bibitem[\protect\citeauthoryear{Kaptelinin}{Kaptelinin}{2003}]%
        {kaptelinin2003umea}
\bibfield{author}{\bibinfo{person}{Victor Kaptelinin}.}
  \bibinfo{year}{2003}\natexlab{}.
\newblock \showarticletitle{UMEA: translating interaction histories into
  project contexts}. In \bibinfo{booktitle}{\emph{Proceedings of the SIGCHI
  conference on Human factors in computing systems}}.
  \bibinfo{pages}{353--360}.
\newblock


\bibitem[\protect\citeauthoryear{Kapur, Glisky, and Wilson}{Kapur
  et~al\mbox{.}}{2004}]%
        {kapurexternal}
\bibfield{author}{\bibinfo{person}{Narinder Kapur},
  \bibinfo{person}{Elizabeth~L Glisky}, {and} \bibinfo{person}{Barbara~A
  Wilson}.} \bibinfo{year}{2004}\natexlab{}.
\newblock \showarticletitle{External memory aids and computers in memory
  rehabilitation}.
\newblock \bibinfo{journal}{\emph{The essential handbook of memory disorders
  for clinicians}} (\bibinfo{year}{2004}), \bibinfo{pages}{301--321}.
\newblock


\bibitem[\protect\citeauthoryear{Kiss, Green, and Kao}{Kiss
  et~al\mbox{.}}{2005}]%
        {kiss2005disease}
\bibfield{author}{\bibinfo{person}{Istvan~Z Kiss}, \bibinfo{person}{Darren~M
  Green}, {and} \bibinfo{person}{Rowland~R Kao}.}
  \bibinfo{year}{2005}\natexlab{}.
\newblock \showarticletitle{Disease contact tracing in random and clustered
  networks}.
\newblock \bibinfo{journal}{\emph{Proceedings of the Royal Society B:
  Biological Sciences}} \bibinfo{volume}{272}, \bibinfo{number}{1570}
  (\bibinfo{year}{2005}), \bibinfo{pages}{1407--1414}.
\newblock


\bibitem[\protect\citeauthoryear{Kosara, Miksch, Shahar, and Johnson}{Kosara
  et~al\mbox{.}}{2002}]%
        {asbru}
\bibfield{author}{\bibinfo{person}{Robert Kosara}, \bibinfo{person}{Silvia
  Miksch}, \bibinfo{person}{Yuval Shahar}, {and} \bibinfo{person}{Peter
  Johnson}.} \bibinfo{year}{2002}\natexlab{}.
\newblock \showarticletitle{AsbruView: capturing complex, time-oriented plans
  Ñ beyond flow charts}.
\newblock In \bibinfo{booktitle}{\emph{Diagrammatic Representation and
  Reasoning}}. \bibinfo{pages}{535--549}.
\newblock


\bibitem[\protect\citeauthoryear{Li, Zheng, Xie, Chen, Liu, and Ma}{Li
  et~al\mbox{.}}{2008}]%
        {li2008mining}
\bibfield{author}{\bibinfo{person}{Quannan Li}, \bibinfo{person}{Yu Zheng},
  \bibinfo{person}{Xing Xie}, \bibinfo{person}{Yukun Chen},
  \bibinfo{person}{Wenyu Liu}, {and} \bibinfo{person}{Wei-Ying Ma}.}
  \bibinfo{year}{2008}\natexlab{}.
\newblock \showarticletitle{Mining user similarity based on location history}.
  In \bibinfo{booktitle}{\emph{Proceedings of the 16th ACM SIGSPATIAL
  conference on Advances in geographic information systems}}. ACM,
  \bibinfo{pages}{34}.
\newblock


\bibitem[\protect\citeauthoryear{Liu and Singh}{Liu and Singh}{2004}]%
        {liu2004conceptnet}
\bibfield{author}{\bibinfo{person}{Hugo Liu} {and} \bibinfo{person}{Push
  Singh}.} \bibinfo{year}{2004}\natexlab{}.
\newblock \showarticletitle{ConceptNet: a practical commonsense reasoning
  tool-kit}.
\newblock \bibinfo{journal}{\emph{BT technology journal}} \bibinfo{volume}{22},
  \bibinfo{number}{4} (\bibinfo{year}{2004}), \bibinfo{pages}{211--226}.
\newblock


\bibitem[\protect\citeauthoryear{Miller}{Miller}{1995}]%
        {miller1995wordnet}
\bibfield{author}{\bibinfo{person}{George~A Miller}.}
  \bibinfo{year}{1995}\natexlab{}.
\newblock \showarticletitle{WordNet: a lexical database for English}.
\newblock \bibinfo{journal}{\emph{Commun. ACM}} \bibinfo{volume}{38},
  \bibinfo{number}{11} (\bibinfo{year}{1995}), \bibinfo{pages}{39--41}.
\newblock


\bibitem[\protect\citeauthoryear{Murakami and Mitsuhashi}{Murakami and
  Mitsuhashi}{2012}]%
        {murakami2012system}
\bibfield{author}{\bibinfo{person}{Harumi Murakami} {and}
  \bibinfo{person}{Kenta Mitsuhashi}.} \bibinfo{year}{2012}\natexlab{}.
\newblock \showarticletitle{A System for Creating User’s Knowledge Space from
  Various Information Usages to Support Human Recollection}.
\newblock \bibinfo{journal}{\emph{International Journal of Advancements in
  Computing Technology}} \bibinfo{volume}{4}, \bibinfo{number}{22}
  (\bibinfo{year}{2012}), \bibinfo{pages}{496--508}.
\newblock


\bibitem[\protect\citeauthoryear{Naeem, Bigham, and Wang}{Naeem
  et~al\mbox{.}}{2007}]%
        {ADLasbru}
\bibfield{author}{\bibinfo{person}{Usman Naeem}, \bibinfo{person}{John Bigham},
  {and} \bibinfo{person}{Jinfu Wang}.} \bibinfo{year}{2007}\natexlab{}.
\newblock \showarticletitle{Recognising activities of daily life using
  hierarchical plans}. In \bibinfo{booktitle}{\emph{European Conf. on Smart
  Sensing and Context}}. \bibinfo{pages}{175--189}.
\newblock


\bibitem[\protect\citeauthoryear{Organization, for Disease~Control, Prevention,
  et~al\mbox{.}}{Organization et~al\mbox{.}}{2015}]%
        {world2015implementation}
\bibfield{author}{\bibinfo{person}{World~Health Organization},
  \bibinfo{person}{Centers for Disease~Control}, \bibinfo{person}{Prevention},
  {et~al\mbox{.}}} \bibinfo{year}{2015}\natexlab{}.
\newblock \bibinfo{booktitle}{\emph{Implementation and management of contact
  tracing for Ebola virus disease: emergency guideline}}.
\newblock \bibinfo{type}{{T}echnical {R}eport}. \bibinfo{institution}{World
  Health Organization}.
\newblock


\bibitem[\protect\citeauthoryear{PEPP-PT}{PEPP-PT}{2020}]%
        {PEPP-PT}
\bibfield{author}{\bibinfo{person}{PEPP-PT}.} \bibinfo{year}{2020}\natexlab{}.
\newblock \showarticletitle{Pan-European Privacy-Preserving Proximity Tracing}.
\newblock  (\bibinfo{year}{2020}).
\newblock
\urldef\tempurl%
\url{https://www.pepp-pt.org/}
\showURL{%
\tempurl}


\bibitem[\protect\citeauthoryear{Raskar, Schunemann, Barbar, Vilcans, Gray,
  Vepakomma, Kapa, Nuzzo, Gupta, Berke, et~al\mbox{.}}{Raskar
  et~al\mbox{.}}{2020}]%
        {raskar2020apps}
\bibfield{author}{\bibinfo{person}{Ramesh Raskar}, \bibinfo{person}{Isabel
  Schunemann}, \bibinfo{person}{Rachel Barbar}, \bibinfo{person}{Kristen
  Vilcans}, \bibinfo{person}{Jim Gray}, \bibinfo{person}{Praneeth Vepakomma},
  \bibinfo{person}{Suraj Kapa}, \bibinfo{person}{Andrea Nuzzo},
  \bibinfo{person}{Rajiv Gupta}, \bibinfo{person}{Alex Berke}, {et~al\mbox{.}}}
  \bibinfo{year}{2020}\natexlab{}.
\newblock \showarticletitle{Apps gone rogue: Maintaining personal privacy in an
  epidemic}.
\newblock \bibinfo{journal}{\emph{arXiv preprint arXiv:2003.08567}}
  (\bibinfo{year}{2020}).
\newblock


\bibitem[\protect\citeauthoryear{Schacter}{Schacter}{2002}]%
        {schacter2002seven}
\bibfield{author}{\bibinfo{person}{Daniel~L Schacter}.}
  \bibinfo{year}{2002}\natexlab{}.
\newblock \bibinfo{booktitle}{\emph{The seven sins of memory: How the mind
  forgets and remembers}}.
\newblock \bibinfo{publisher}{HMH}.
\newblock


\bibitem[\protect\citeauthoryear{Schank and Abelson}{Schank and
  Abelson}{1977}]%
        {schank1977}
\bibfield{author}{\bibinfo{person}{R Schank} {and} \bibinfo{person}{R
  Abelson}.} \bibinfo{year}{1977}\natexlab{}.
\newblock \showarticletitle{Scripts, plans, goals, and understanding: an
  inquiry into human knowledge structures.}
\newblock  (\bibinfo{year}{1977}).
\newblock


\bibitem[\protect\citeauthoryear{Sellen and Whittaker}{Sellen and
  Whittaker}{2010}]%
        {sellen2010beyond}
\bibfield{author}{\bibinfo{person}{A. Sellen} {and} \bibinfo{person}{S.
  Whittaker}.} \bibinfo{year}{2010}\natexlab{}.
\newblock \showarticletitle{Beyond total capture: a constructive critique of
  lifelogging}.
\newblock \bibinfo{journal}{\emph{Communic. of the ACM}}
  (\bibinfo{year}{2010}).
\newblock


\bibitem[\protect\citeauthoryear{Shafer}{Shafer}{1986}]%
        {shafer1986combination}
\bibfield{author}{\bibinfo{person}{Glenn Shafer}.}
  \bibinfo{year}{1986}\natexlab{}.
\newblock \showarticletitle{The combination of evidence}.
\newblock \bibinfo{journal}{\emph{International Journal of Intelligent
  Systems}} \bibinfo{volume}{1}, \bibinfo{number}{3} (\bibinfo{year}{1986}),
  \bibinfo{pages}{155--179}.
\newblock


\bibitem[\protect\citeauthoryear{Snyder and Kanich}{Snyder and Kanich}{2013}]%
        {snyder2013cloudsweeper}
\bibfield{author}{\bibinfo{person}{Peter Snyder} {and} \bibinfo{person}{Chris
  Kanich}.} \bibinfo{year}{2013}\natexlab{}.
\newblock \showarticletitle{Cloudsweeper: enabling data-centric document
  management for secure cloud archives}. In
  \bibinfo{booktitle}{\emph{Proceedings of the 2013 ACM workshop on Cloud
  computing security workshop}}. \bibinfo{pages}{47--54}.
\newblock


\bibitem[\protect\citeauthoryear{Stelmach}{Stelmach}{[n.d.]}]%
        {natty}
\bibfield{author}{\bibinfo{person}{Stelmach}.}
  \bibinfo{year}{[n.d.]}\natexlab{}.
\newblock \bibinfo{title}{Natty Date Parser}.
\newblock \bibinfo{howpublished}{\url{http://natty.joestelmach.com}}.
\newblock
\newblock
\shownote{Accessed: 2019.}


\bibitem[\protect\citeauthoryear{Thakur}{Thakur}{2016}]%
        {thakur2016personalization}
\bibfield{author}{\bibinfo{person}{Shashi Thakur}.}
  \bibinfo{year}{2016}\natexlab{}.
\newblock \showarticletitle{Personalization for google now: user understanding
  and application to information recommendation and exploration}. In
  \bibinfo{booktitle}{\emph{Proceedings of the 10th ACM Conference on
  Recommender Systems}}. \bibinfo{pages}{3--3}.
\newblock


\bibitem[\protect\citeauthoryear{Tulving and Donaldson}{Tulving and
  Donaldson}{1972}]%
        {tulving1972organisation}
\bibfield{author}{\bibinfo{person}{E Tulving} {and} \bibinfo{person}{W
  Donaldson}.} \bibinfo{year}{1972}\natexlab{}.
\newblock \bibinfo{title}{Organisation of Memory. New York and London}.
\newblock
\newblock


\bibitem[\protect\citeauthoryear{Van Den~Hoven, Sas, and Whittaker}{Van
  Den~Hoven et~al\mbox{.}}{2012}]%
        {van2012introduction}
\bibfield{author}{\bibinfo{person}{Elise Van Den~Hoven},
  \bibinfo{person}{Corina Sas}, {and} \bibinfo{person}{Steve Whittaker}.}
  \bibinfo{year}{2012}\natexlab{}.
\newblock \showarticletitle{Introduction to this special issue on designing for
  personal memories: past, present, and future}.
\newblock \bibinfo{journal}{\emph{Human--Computer Interaction}}
  \bibinfo{volume}{27}, \bibinfo{number}{1-2} (\bibinfo{year}{2012}),
  \bibinfo{pages}{1--12}.
\newblock


\bibitem[\protect\citeauthoryear{Vianna, Yong, Xia, Marian, and Nguyen}{Vianna
  et~al\mbox{.}}{2014}]%
        {iiweb}
\bibfield{author}{\bibinfo{person}{Daniela Vianna},
  \bibinfo{person}{Alicia-Michelle Yong}, \bibinfo{person}{Chaolun Xia},
  \bibinfo{person}{Am\'elie Marian}, {and} \bibinfo{person}{Thu~D. Nguyen}.}
  \bibinfo{year}{2014}\natexlab{}.
\newblock \showarticletitle{A Tool for Personal Data Extraction}. In
  \bibinfo{booktitle}{\emph{Proceedings of the 10th International Workshop on
  Information Integration on the Web (IIWeb 2014)}}.
\newblock


\bibitem[\protect\citeauthoryear{Whittaker, Bellotti, and Gwizdka}{Whittaker
  et~al\mbox{.}}{2006}]%
        {whittaker2006email}
\bibfield{author}{\bibinfo{person}{Steve Whittaker}, \bibinfo{person}{Victoria
  Bellotti}, {and} \bibinfo{person}{Jacek Gwizdka}.}
  \bibinfo{year}{2006}\natexlab{}.
\newblock \showarticletitle{Email in personal information management}.
\newblock \bibinfo{journal}{\emph{Commun. ACM}} \bibinfo{volume}{49},
  \bibinfo{number}{1} (\bibinfo{year}{2006}), \bibinfo{pages}{68--73}.
\newblock


\bibitem[\protect\citeauthoryear{Whittaker and Sidner}{Whittaker and
  Sidner}{1996}]%
        {whittaker1996email}
\bibfield{author}{\bibinfo{person}{Steve Whittaker} {and}
  \bibinfo{person}{Candace Sidner}.} \bibinfo{year}{1996}\natexlab{}.
\newblock \showarticletitle{Email overload: exploring personal information
  management of email}. In \bibinfo{booktitle}{\emph{Proceedings of the SIGCHI
  conference on Human factors in computing systems}}.
  \bibinfo{pages}{276--283}.
\newblock


\end{thebibliography}

\end{document}